\documentclass[aps,pre,longbibliography]{revtex4-1}
\usepackage{color}
\usepackage{amssymb}
\usepackage{amsmath,bm}
\usepackage{graphicx}

\begin{document}
\title{The three-dimensional turbulent cellular flow}

\author{S. Berti}
\affiliation{Univ. Lille, ULR 7512, Unit\'e de M\'ecanique de Lille Joseph
Boussinesq (UML), 59000 Lille, France}

\author{G. Boffetta}\thanks{\emph{Corresponding author:} guido.boffetta@unito.it}
\affiliation{Dipartimento di Fisica and INFN, Universit\`a degli Studi di
Torino, via P. Giuria 1, 10125 Torino, Italy.}

\author{S. Musacchio}
\affiliation{Dipartimento di Fisica and INFN, Universit\`a degli Studi di
Torino, via P. Giuria 1, 10125 Torino, Italy.}

\begin{abstract}
We study, by means of extensive direct numerical simulations, the turbulent
flow produced by a two-dimensional cellular forcing in a cubic box with
periodic boundary conditions.  In spite of the strong anisotropy of the
forcing, we find that turbulence recovers almost complete isotropy at small
scales.  Nonetheless, the signature of the forcing remains in the mean flow
(averaged over time and over the homogeneous direction) and this allows to
introduce a friction factor,  whose dependence on the Reynolds number is
investigated.  We further find that the flow is characterized by large temporal
fluctuations of the total energy, as a consequence of the exchange between the
forced mean flow at large scales and turbulent fluctuations at small scales.
Such temporal fluctuations produce a correction to the energy spectrum that can
be explained by a simple dimensional argument.
\end{abstract}


\maketitle

\section{Introduction}

Theoretical and numerical studies of turbulent flows can be divided into two
broad categories. The first class of studies, mainly motivated by 
experiments and practical applications, considers turbulence as generated 
by the interaction of the flow with a solid object. 
The most studied example is the case of turbulence produced at the boundaries
of a container, as in channel and pipe flows or in Taylor-Couette flow 
\cite{avila2022transition,grossmann2016high}. 
The second category focuses mainly on intrinsic, bulk properties of 
turbulence far from boundaries, which may be expected to give 
universal statistics independently 
of the way the flow is generated. These studies are usually based on the 
simplest possible geometry in the absence of boundaries, the so-called 
homogeneous-isotropic turbulence 
in periodic domains.

Between these two widely studied flows, there is space for another
group of inhomogeneous and anisotropic flows in the absence of boundaries.
In this case homogeneity and isotropy are broken not by physical boundaries, 
but by the body force that sustains the flow. 
An important class in this group is that of forces able to produce 
a stationary laminar flow, balanced by the viscous term, 
which is a solution of the Navier-Stokes equation.
The most studied examples in this class are the Kolmogorov flow 
\cite{meshalkin1961investigation,musacchio2014turbulent}, 
the Taylor-Green vortex \cite{taylor1937mechanism,brachet1983small} and the
Arnold-Beltrami-Childress flow \cite{arnold1965topologie}.

In this work we consider another important 
instance in this class, the cellular flow
produced by a two-dimensional, two-component periodic forcing
that does not depend on the third (vertical) dimension. 
At low Reynolds ($Re$) number, the laminar flow consists of a regular array of 
counter-rotating vortices. 
At high $Re$ values, the laminar solution is unstable and the 
flow produced by the cellular forcing becomes turbulent. 
We investigate the statistical properties of the velocity field, 
focusing on the recovery of isotropy at small scales and on the relevance of 
velocity fluctuations in the direction normal to the plane of the forcing. 
In addition, we characterize the properties of the mean flow. 
The presence of a well defined mean velocity field, which essentially inherits 
its spatial structure from the forcing, then allows us to define a 
friction factor based on the vorticity budget,
equivalent to the friction factor based on
the momentum budget in the case of the Kolmogorov 
flow~\cite{musacchio2014turbulent}. 
As in wall-bounded flows,
the friction factor quantifies the (inverse) efficiency of the work done
by the external force to sustain the mean flow.
Finally, we discuss the impact of the temporal behavior of the turbulent
dynamics on the flow statistics.  In particular, we find strong fluctuations of
the mean-flow intensity, which correspond to an alternation of phases in which
the flow displays dominant two-dimensional (2D) and three-dimensional (3D)
structures~\cite{goto2017hierarchy}.  The latter temporal fluctuations induce a
correction to the kinetic energy spectrum, whose form can be explained by a
dimensional argument.

Besides its theoretical interest, which is due to the simple structure of the
mean flow, the cellular flow also represents a useful idealized model for
geophysical applications~\cite{majda2006nonlinear,forgia2022numerical}.  For
instance, the present configuration may have some relevance for ocean
turbulence, where large (mesoscale) coherent eddies, characterized by
essentially horizontal motions, populate the flow together with smaller
(submesoscale) turbulent features displaying much more important vertical
velocities. In spite of the idealized nature of the cellular flow in comparison
with realistic oceanographic conditions, its settings may allow to gain insight
into the basic mechanisms controlling the effect of turbulence on the transport
of inert or chemical and biological tracers in ocean flows.

This article is organized as follows. In Sec.~\ref{sec:model} we introduce the
model and the simulation settings. Section~\ref{sec:statistics} discusses the
results about the statistical properties of turbulence in this flow. There, we
also separately consider the main features of the mean vorticity field
(Sec.~\ref{sec:meanflow}) and the role of temporal fluctuations
(Sec.~\ref{sec:fluct}). 
Finally, discussions and conclusions are presented in
Sec.~\ref{sec:concl}.

\section{Model and numerical simulations}
\label{sec:model}

\begin{figure}[h!]
\centerline{
\includegraphics[scale=0.30]{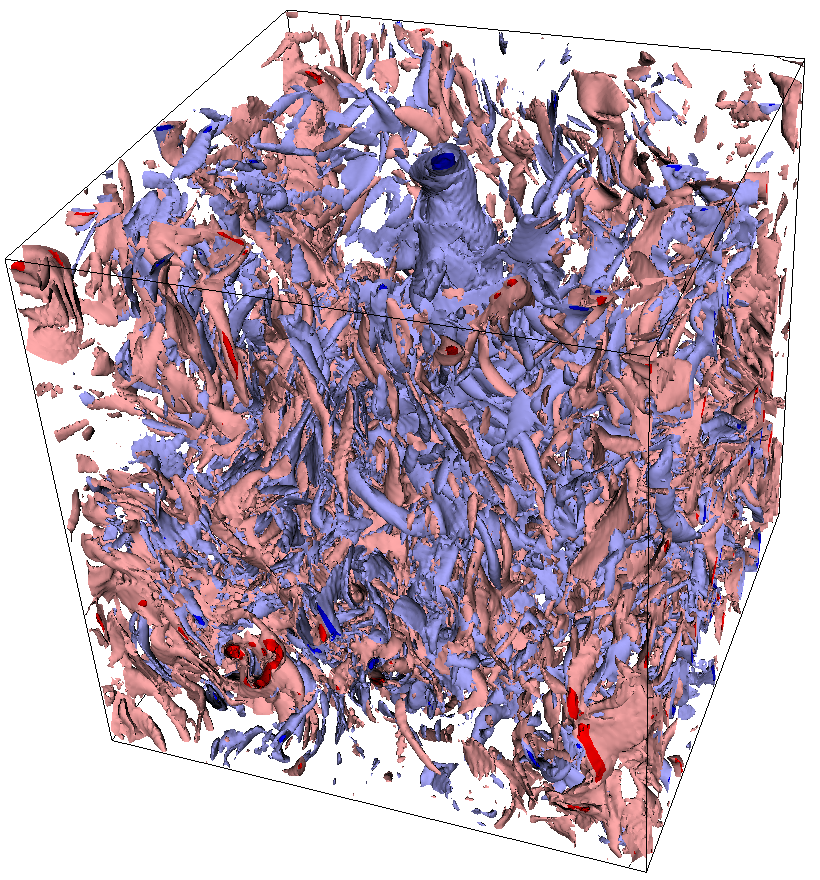}
\includegraphics[scale=0.30]{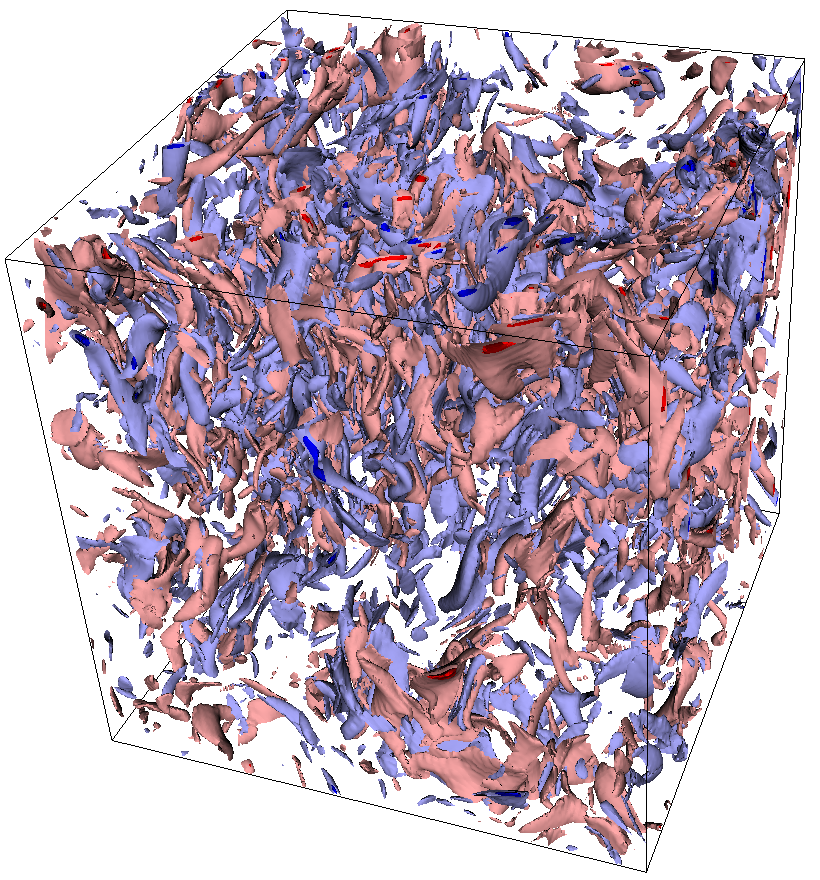}}
\caption{Snapshots of the vertical vorticity field for the simulation 
with $F=0.04$ at two different times ($t=200$ on the left, and $t=800$ on the right) 
corresponding to a maximum and a minimum of kinetic energy.
Blue (red) regions correspond to positive (negative) values of 
$\omega_z({\bm x},t)$.}
\label{fig1}
\end{figure}

We consider the 3D Navier-Stokes (NS) equation
for an incompressible velocity field ${\bm u}({\bm x},t)$,
written for the vorticity field ${\bm \omega}={\bm \nabla} \times {\bm u}$ as
\begin{equation}
\partial_t {\bm \omega} + {\bm u \cdot \nabla \omega} = 
{\bm \omega} \cdot {\bm \nabla u} + \nu \nabla^2 {\bm \omega} + {\bm \nabla} \times {\bm f}_u
\label{eq1}
\end{equation}
where $\nu$ is the kinematic viscosity and
${\bm f}_u=(-F \sin(K y),F \sin(K x),0)$
is the 2D, two-components body force. 
The NS equation admits in this case the laminar steady solution
for the velocity ${\bm u}_0=(-U_0 \sin(K y), U_0 \sin(K x),0)$
and vorticity ${\bm \omega}_0=(0,0,U_0 K [\cos(K x) + \cos(K y)])$
with $U_0=F/(\nu K^2)$.
This solution is called {\it cellular flow}, 
because it has the form of a regular array of 
vortices with alternating circulation sign.
By rotating the reference frame 
of an angle $\theta= \pi/4$ around the $z$ axis, 
i.e. with the coordinate transformation
$x' = (y+x)\sqrt{2}/2$, $y' = (y-x)\sqrt{2}/2$, $z'=z$, 
and by rescaling the wavenumber as $K'=K\sqrt{2}/2$,  
the cellular flow can be written in the equivalent form 
${\bm \omega}_0(x',y',z')=(0,0,2\sqrt{2} U_0 K' [\cos(K' x')\cos(K' y')])$,  
which corresponds to the 2D stationary Taylor-Green 
vortex~\cite{taylor1937mechanism}. 
The laminar solution of the cellular flow is linearly stable when 
$Re=U_0/(\nu K) \le Re_c = 2 \sqrt{2}$~\cite{sivashinsky1985negative}.
For $Re>Re_c$ the basic flow develops large scale perturbations which,
increasing the Reynolds number, become chaotic 
\cite{perlekar2010turbulence} and eventually, for 
$Re \gg Re_c$, the flow becomes 3D and turbulent. 

In the following we consider the situation $Re \gg Re_c$ for which 
no analytic approach is available and we therefore exploit direct
high-resolution numerical simulations of the NS equations.
To this aim, we use a fully-dealiased, pseudo-spectral 
code that integrates Eq.~(\ref{eq1}) on a periodic cubic 
domain of side $L=2\pi$ at resolution $N^3$. 
We fixed the forcing wavenumber to $K=1$ and we the viscosity to 
$\nu=10^{-3}$, so that the only control parameter is the forcing amplitude $F$.
For each value of $F$, we perform a bootstrap simulation to reach
a statistically stationary state, followed by very long simulations 
(over $25-100$ eddy-turnover-times, depending on $F$) 
during which average quantities are computed. 
Table~\ref{table1} reports the main parameters of the 
simulations. 

\begin{table}[h!]
\begin{tabular}{cccccccccc}
$F$ & $E$ & $u_{x}$ & $u_{z}$ & $U$ & $Re$ & $\varepsilon$ & $\eta$ & $\tau_{\eta}$ & $T$ \\ \hline
$0.005$ & $0.032$ & $0.16$ & $0.11$ & $0.08$ & $80$ & $8.3 \times 10^{-4}$ & $3.3 \times 10^{-2}$ & $1.1$ & $39$ \\
$0.01$ & $0.071$ & $0.24$ & $0.15$ & $0.12$ & $124$ & $2.5 \times 10^{-3}$ & $2.5 \times 10^{-2}$ & $0.64$ & $28$ \\
$0.02$ & $0.16$ & $0.36$ & $0.22$ & $0.19$ & $188$ & $7.5 \times 10^{-3}$ & $1.9 \times 10^{-2}$ & $0.37$ & $21$ \\
$0.04$ & $0.29$ & $0.49$ & $0.31$ & $0.25$ & $252$ & $2.0 \times 10^{-2}$ & $1.5 \times 10^{-2}$ & $0.22$ & $15$ \\
$0.08$ & $0.57$ & $0.69$ & $0.44$ & $0.34$ & $343$ & $5.5 \times 10^{-2}$ & $1.2 \times 10^{-2}$ & $0.13$ & $10$ \\
$0.16$ & $1.24$ & $1.01$ & $0.65$ & $0.51$ & $508$ & $0.166$ & $8.8 \times 10^{-3}$ & $0.077$ & $7.5$ \\
$0.32$ & $2.36$ & $1.39$ & $0.90$ & $0.69$ & $692$ & $0.450$ & $6.9 \times 10^{-3}$ & $0.047$ & $5.2$
\end{tabular}
\caption{Parameters of the simulations: $F$ is the forcing amplitude, 
$E$ the mean kinetic energy, 
$u_{x}$ and $u_{z}$ the rms value of the 
$x$ and $z$ velocity components, respectively, 
$U$ the amplitude of mode $(k_x,k_y)=(0,1)$ of $\overline{u_x}(x,y)$ (see text),
$Re=U/(K \nu)$ the Reynolds number,
$\varepsilon=\nu \langle ({\bm \nabla} {\bm u})^2 \rangle$
the mean energy dissipation, $\eta=(\nu^3/\varepsilon)^{1/4}$ the Kolmogorov scale,
$\tau_{\eta}=(\nu/\varepsilon)^{1/2}$ the Kolmogorov timescale, 
and $T=E/\varepsilon$ the integral timescale. 
The forcing wavenumber $K=1$ and the viscosity $\nu=10^{-3}$ are 
fixed for all simulations. The spatial 
resolution is $N=256$ for runs up to $F=0.04$, and $N=512$ for larger $F$. 
For all simulations $k_{max} \eta \ge 1$.}
\label{table1}
\end{table}

Figure~\ref{fig1} shows two examples of the $z$ component of the vorticity
field for the run at $F=0.04$ at two different instants of time, corresponding to
a maximum (left) and a minimum (right) of the global kinetic energy
(see Fig.~\ref{fig2}). It is evident that the flow is fully
turbulent with positive and negative vorticity structures, which appear
elongated mostly along the homogeneous direction $z$, in 
particular for the high energy case (left panel in Fig.~\ref{fig1}), 
where a large scale vortex at 
the center of the computational box is observable. 


\section{Statistics of the turbulent flow}
\label{sec:statistics}

\begin{figure}[h!]
\centerline{\includegraphics[scale=0.75]{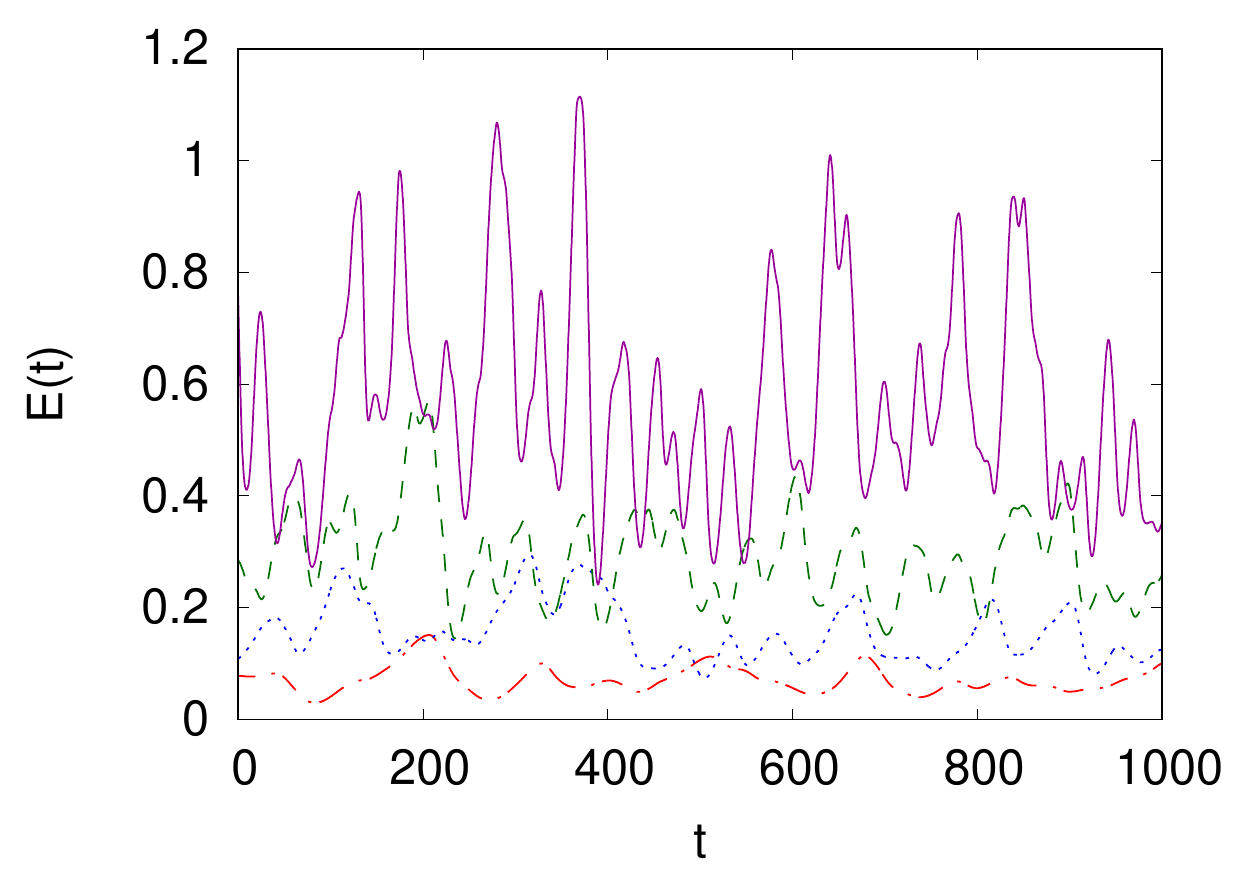}
\includegraphics[scale=0.75]{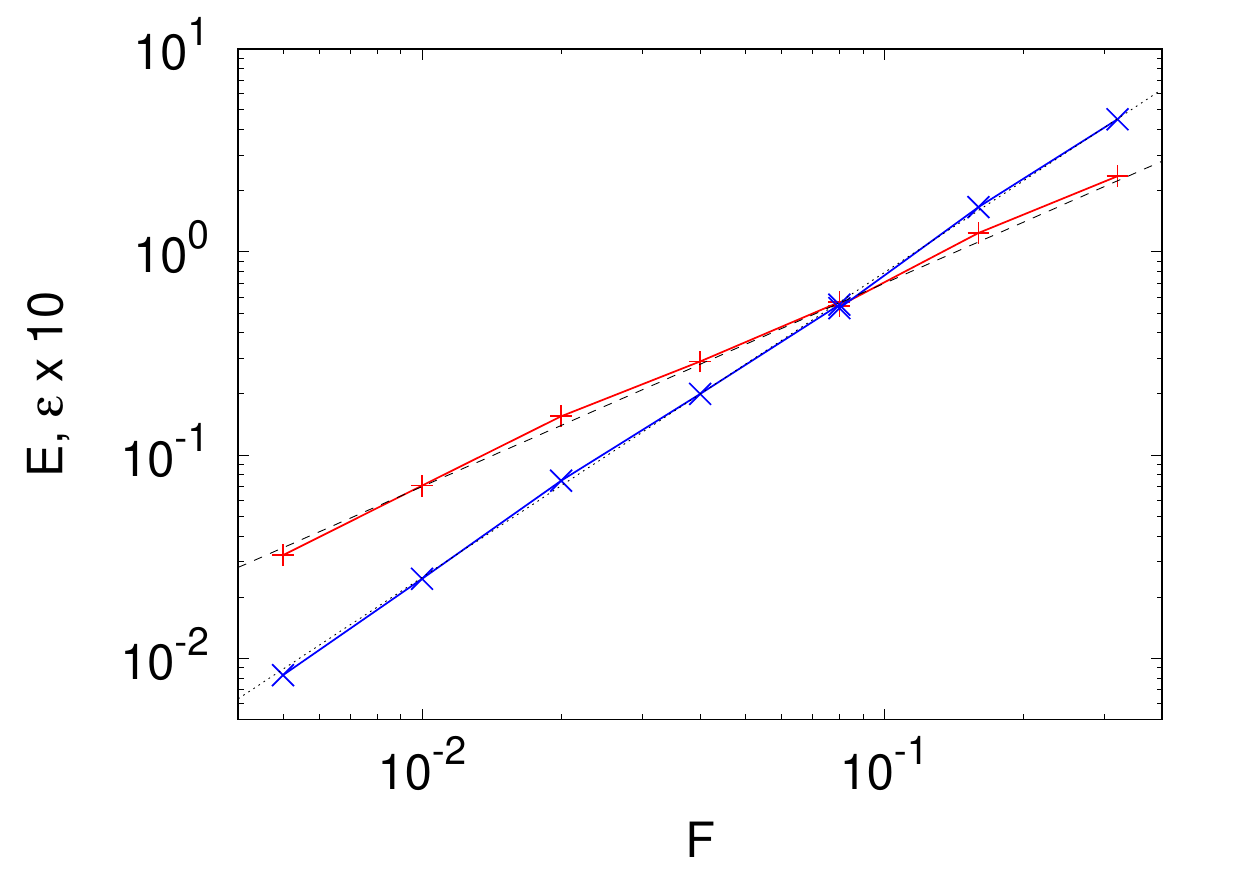}}
\caption{
Left panel: Kinetic energy $E(t)$ as a function of time for simulations
with $F=0.01$ (red dot-dashed line), $F=0.02$ (blue dotted line),
$F=0.04$ (green dashed line) and $F=0.08$ (purple continuous line).
Right panel: Time-averaged kinetic energy (red line, plus symbols) and 
energy dissipation rate (blue line, cross symbols) as a 
function of the forcing amplitude $F$. 
The dashed (dotted) line has slope $1$ ($3/2$).
The energy dissipation rate has been rescaled by a factor $10$ for clarity.
}
\label{fig2}
\end{figure}

Figure~\ref{fig2} (left panel) 
shows the time evolution of the mean kinetic energy 
$E = \frac{1}{2} \langle |{\bm u}|^2 \rangle$, where 
$\langle \cdots \rangle = 1/L^3 \int \cdots d^3 x$ 
is the average over the whole domain,   
for a set of simulations in statistically stationary conditions. 
We observe large fluctuations in the instantaneous value of the energy,
with characteristic time proportional to the integral 
timescale $T=E/\varepsilon$ of the flow. 
Because of  these oscillations, very long simulations 
are required in order to reach the convergence of the 
statistical quantities. 

The time-averaged values of the energy 
$E$ and of the energy dissipation rate
$\varepsilon = \nu \langle  ({\bm \nabla} {\bm u})^2 \rangle$ 
increase with the forcing amplitude $F$, 
as shown in Figure~\ref{fig2} (right panel). 
Scaling predictions for $E$ and $\varepsilon$ can be 
derived by simple dimensional arguments. 
Recalling that in stationary conditions the dissipation rate is equal 
to the energy injection rate of the deterministic forcing 
$\varepsilon= \langle {\bm f}_u \cdot {\bm u} \rangle \simeq F E^{1/2}$ 
and it is also equal to the energy flux of the turbulent cascade, 
which can be written on dimensional ground as 
$\varepsilon \simeq E^{3/2}K$, 
we obtain the scaling predictions $E \propto F$ and $\varepsilon \propto F^{3/2}$. 
Both these scaling laws are observed in our simulations 
(see Fig.~\ref{fig2}, right panel). 

\begin{figure}[h!]
\centerline{\includegraphics[scale=0.75]{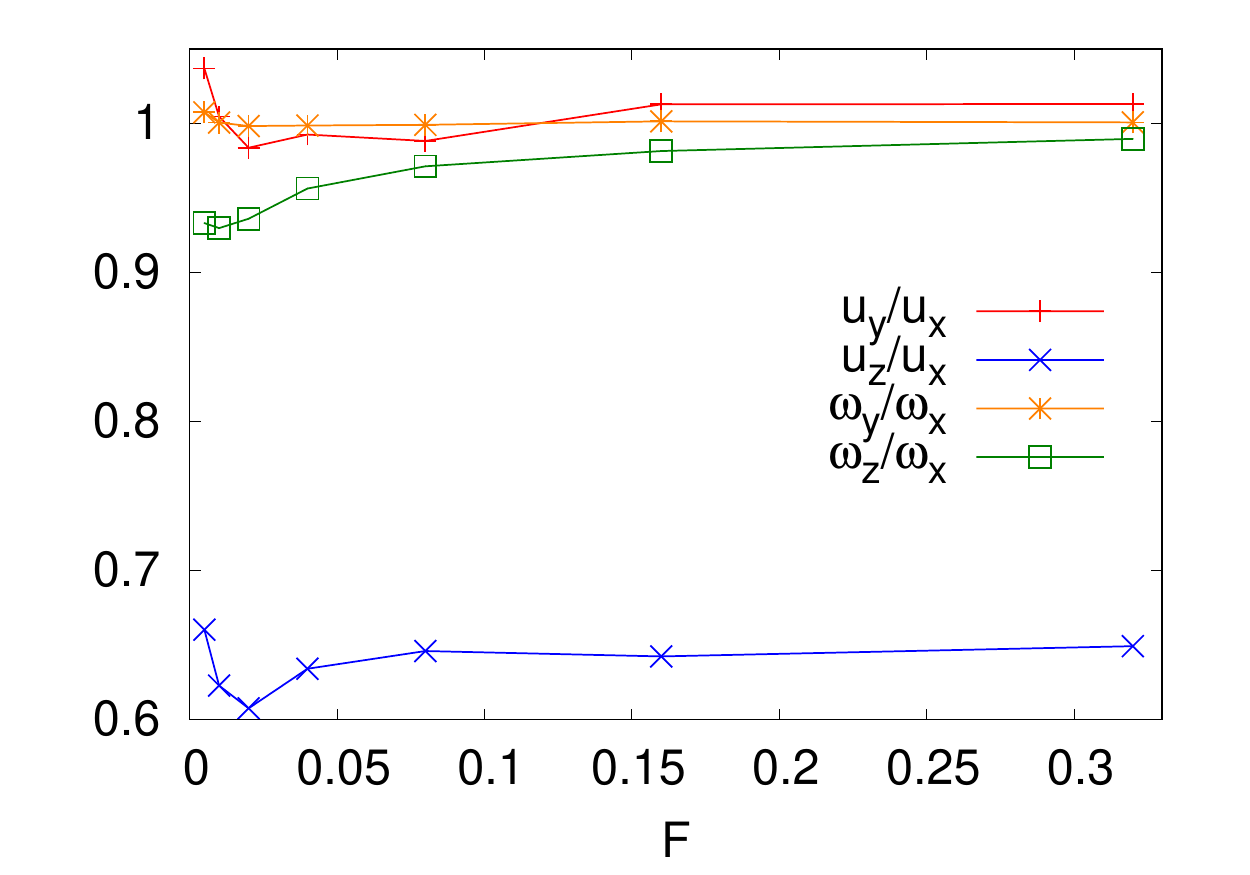}}
\caption{Ratio of the rms values of velocity 
and vorticity components as a function of the forcing amplitude
$F$.}
\label{fig3}
\end{figure}

At large scales, the turbulent cellular flow is anisotropic, 
as a consequence of the forcing, which acts on the 
horizontal components of the velocity only. 
This is shown by the ratio of the vertical to the horizontal root-mean-square 
(rms) velocity 
$u_{z}/u_{x}$, which is close to $0.65$  for all the values 
of the forcing amplitude $F$ (see Fig.~\ref{fig3} ). 
Nonetheless, the ratio of the vertical to the horizontal rms vorticity 
$\omega_{z}/\omega_{x}$ (which is dominated by small-scale contributions)
remains close to unity, revealing that isotropy is recovered at small scales. 
The statistical isotropy of the flow in the $(x,y)$ plane 
is shown by the ratios $u_{y}/u_{x}$ and $\omega_{y}/\omega_{x}$ 
which are close to unity for all the values of $F$. 

\begin{figure}[h!]
\centerline{
\includegraphics[scale=0.75]{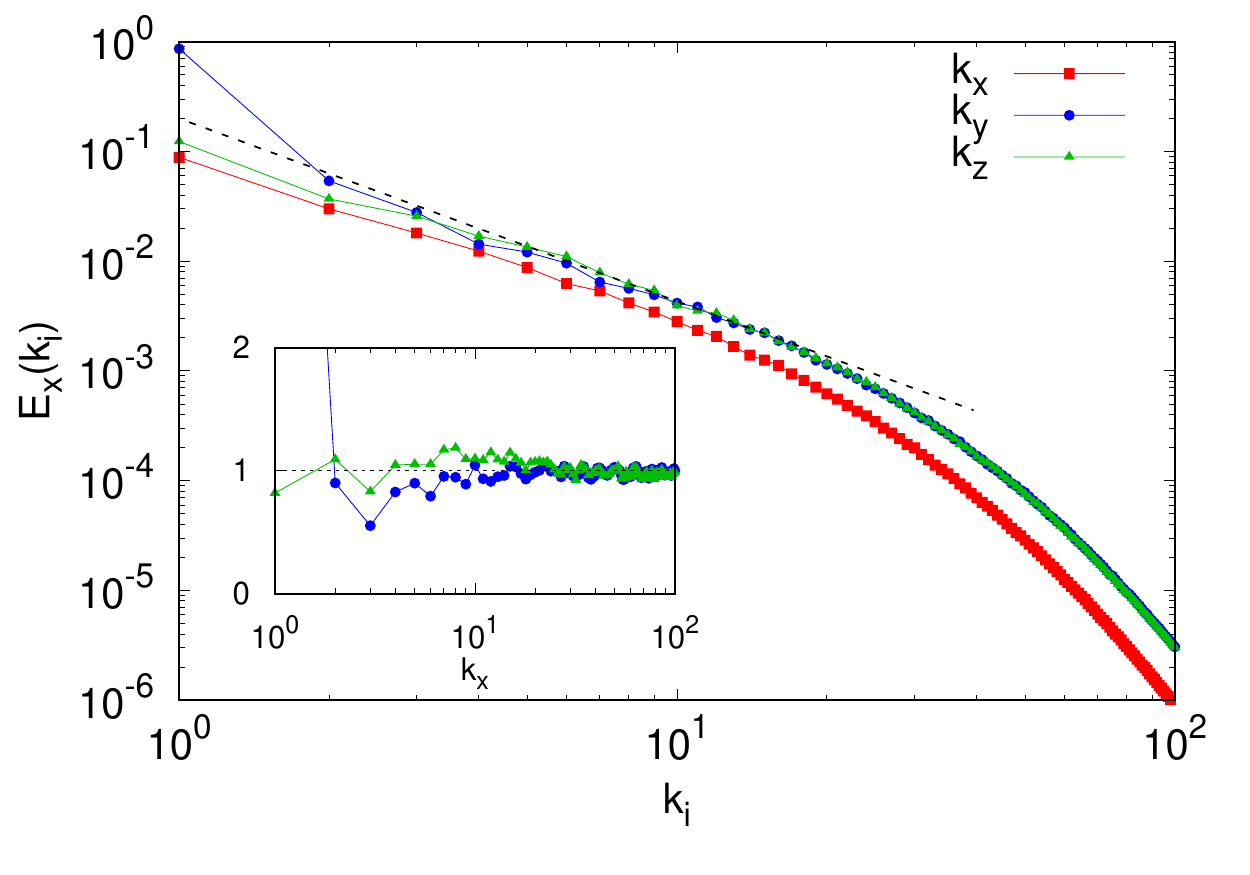}
\includegraphics[scale=0.75]{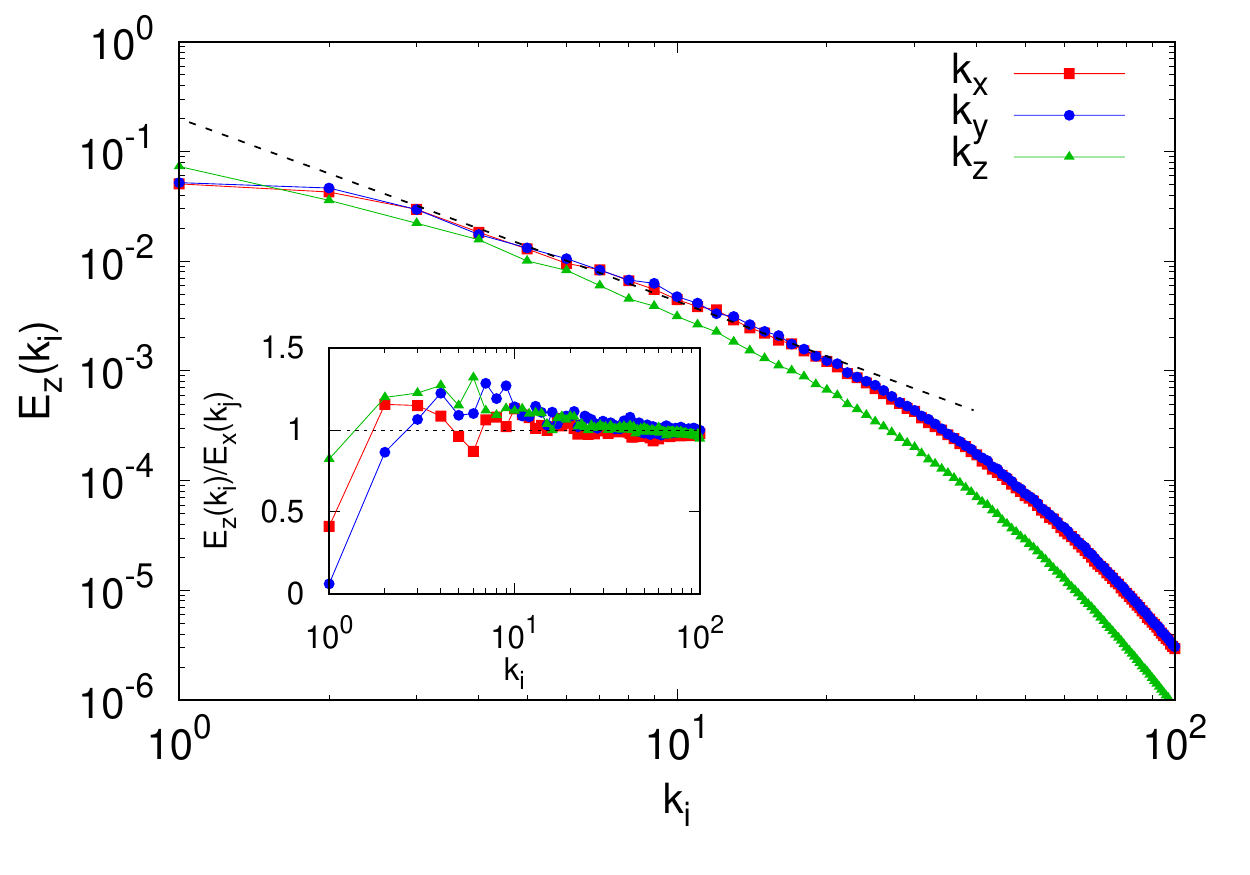}
}
\caption{
One-dimensional energy spectrum for the $x$ (left panel) and $z$ (right panel) 
components of the velocity field as a function of the wavenumbers 
$k_x$ (red squares), $k_y$ (blue circles),  $k_z$ (green triangles), 
for the run with $F=0.32$. 
The dashed line represents the Kolmogorov scaling 
$E(k) \sim \varepsilon^{2/3} k^{-5/3}$.
Inset, left:
ratios $E_y(k_x)/\left[E_x(k_x)/2 - k_x \partial_{k_x} E_x(k_x)/2\right]$ (blue circles)
and $E_z(k_x)/\left[E_x(k_x)/2 - k_x \partial_{k_x} E_x(k_x)/2\right]$ (green triangles).
Inset, right:
ratios $E_z(k_x)/E_x(k_z)$ (red squares), 
$E_z(k_y)/E_x(k_y)$ (blue circles)
and $E_z(k_z)/E_x(k_x)$ (green triangles).
}
\label{fig4}
\end{figure}

The recovery of isotropy at small scales is confirmed by the 
instantaneous one-dimensional (1D) energy spectra 
of the $x$ and $z$ components of the velocity field, 
which are shown in Fig.~\ref{fig4}. 
The 1D energy spectra are computed by taking the 
Fourier transform of the velocity field at a given time in a single spatial direction, 
and then averaged in the other two directions. 
The spectra of the $y$ component of the velocity (not shown) 
are indistinguishable from that of the $x$ component (exchanging $k_x$ with $k_y$). 

At large scale, the 1D spectrum of horizontal velocity $E_x(k_y)$ 
reveals the presence of strong, anisotropic flow at $k_y=1$, 
which is absent in the vertical component $E_z(k_y)$.  
Such energy peak is observed in the transverse horizontal mode, 
i.e. $k_y=1$ for $E_x$ and $k_x=1$ for $E_y$. 
This reflects the vortical structure of the large-scale flow.
Nonetheless, the flow recovers almost complete isotropy at small scales
(i.e., at wavenumbers $k \gtrsim 10$) 
with values of the ratios $E_z(k_x)/E_x(k_z)$, $E_z(k_y)/E_x(k_y)$ and $E_z(k_z)/E_x(k_x)$
close to $1$ (Figure~\ref{fig4}, right inset).

At intermediate scales, the turbulent cascade develops a Kolmogorov spectrum 
$E(k) = C \varepsilon^{2/3} k^{-5/3}$ for both the components 
$E_x$ and $E_z$. 
From the  K\'arm\'an relation~\cite{batchelor1953theory} 
between the second-order transverse and longitudinal structure functions,
which holds under the hypotheses of statistical homongeneity and isotropy, 
it is possible to derive a prediction for the 1D spectra in the transverse and longitudinal modes 
$E_{\perp}(k) = \left[ E_{||}(k) - k \partial_k E_{||}(k) \right] /2$. 
Our results show that the transvese spectra $E_y(k_x)$ and $E_z(k_x)$ 
are in good agreement with the prediction 
$E_{y,z}(k_x) = \left[ E_x((k_x) -  k_x \partial_{k_x} E_x(k_x) \right] /2$ 
at high wavenumbers (see inset of Figure~\ref{fig4}, left panel). 
This is a further confirmation of the recovery of isotropy at small scales.  


\subsection{Mean vorticity field}
\label{sec:meanflow}

\begin{figure}[h!]
\centerline{\includegraphics[scale=0.6]{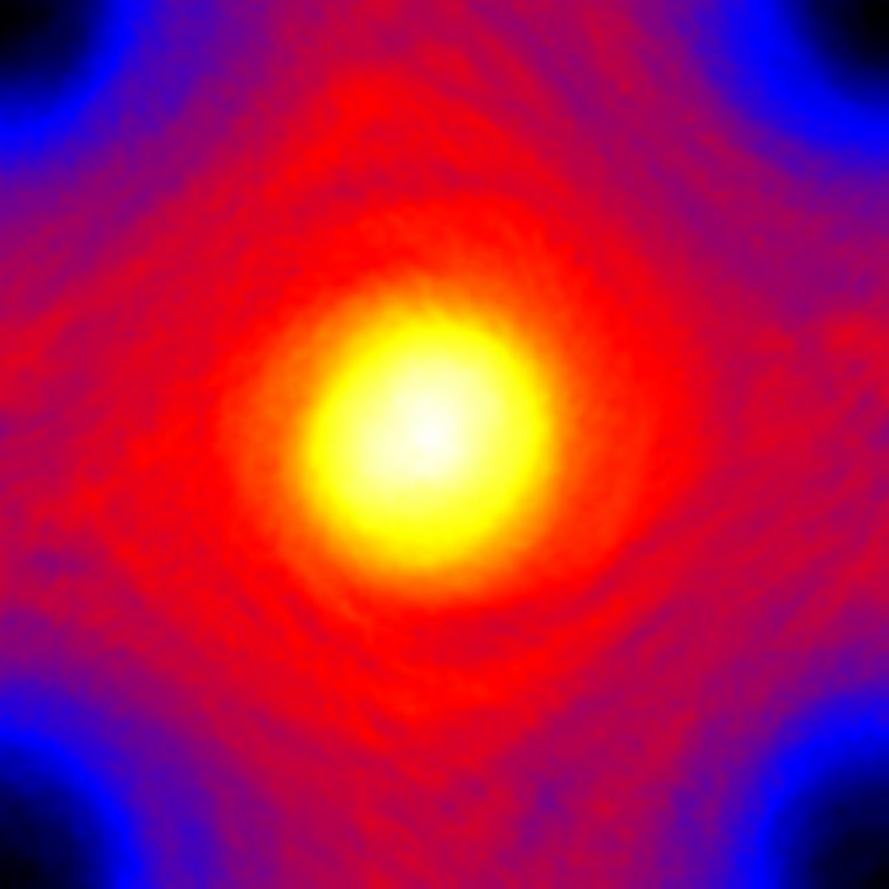}
\includegraphics[scale=0.6]{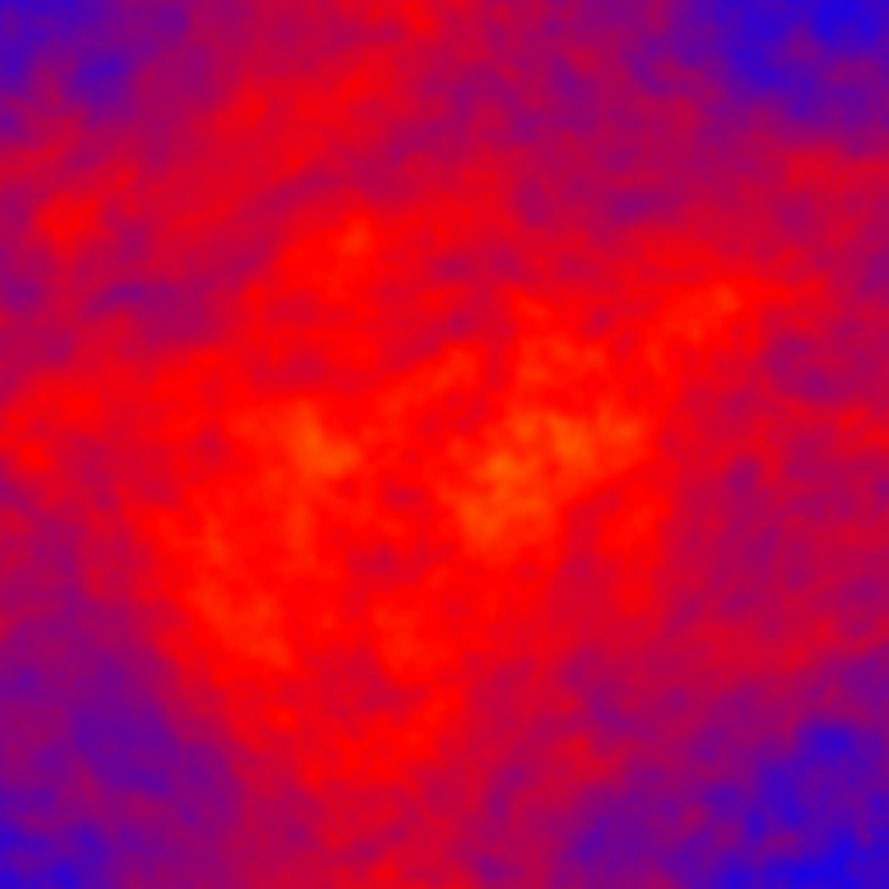}
\includegraphics[scale=0.6]{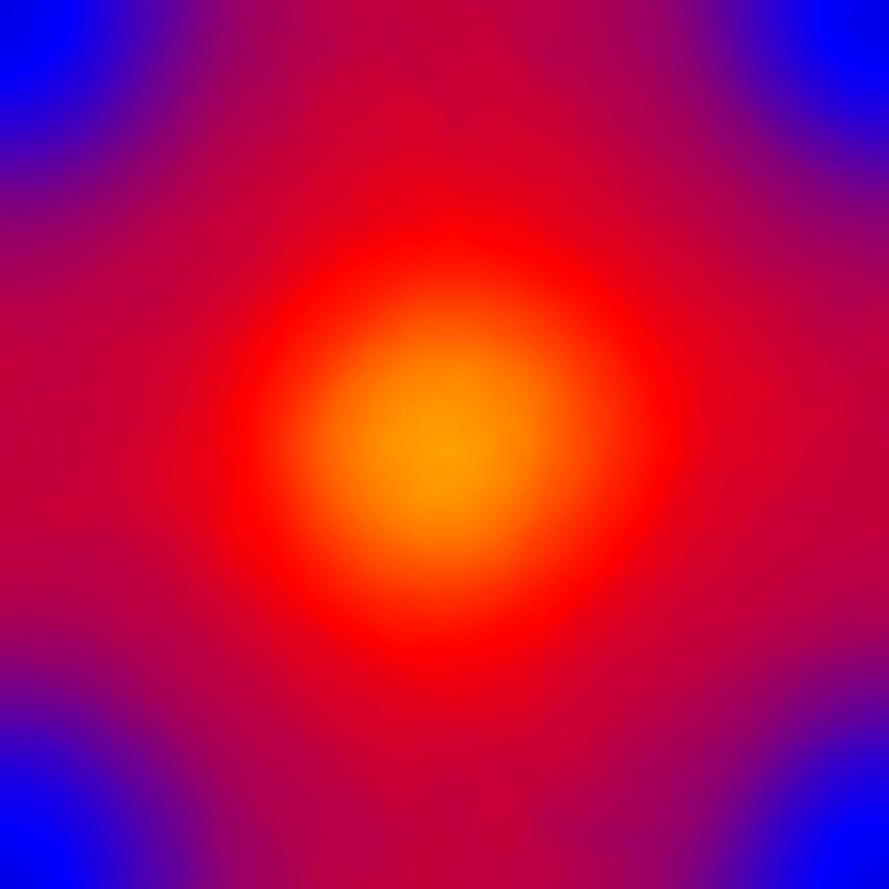}}
\caption{Snapshots of the vertical vorticity $\overline{\omega_z}(x,y)$ 
averaged over the homogeneous $z$ direction, from the run at $F=0.04$.
Black (white) corresponds to maximally negative (positive) vorticity.
Left (center): time $t=200$ ($t=240$) corresponding to a maximum 
(minimum) of the kinetic energy.
Right: temporal average.
}
\label{fig5}
\end{figure}

In spite of the complexity of the vorticity field shown in 
Fig.~\ref{fig1}, it is remarkable
that it is possible to recover the structure of the inhomogeneous forcing
by computing the average of the fields over the homogeneous direction $z$,
which we denote as $\overline{(...)}=(1/L) \int_0^L (...) dz$.
Figure~\ref{fig5} shows that the vertically averaged vorticity field 
$\overline{\omega_z}(x,y)$ displays a clear signature of the forcing.
We also observe the presence of strong temporal fluctuations in the 
instantaneous fields $\overline{\omega_z}$ computed at different times. 
The fluctuations of $\overline{\omega_z}$ are correlated 
with those of the energy, observed in Figure~\ref{fig2}. 
Maxima of the energy occur when the flow displays a strong 2D vortical structure,
while weak 2D structures correspond to the minima of energy. 
The quasi-periodic alternation of these phases 
had been already noted in a previous numerical study~\cite{goto2017hierarchy}, 
and is a consequence of the cellular forcing,
which accumulates energy in the 2D large-scale flow,  
until the latter is discharged 
through the turbulent cascade, which is an intrinsically 
3D process. 

\begin{figure}[h!]
\centerline{\includegraphics[scale=0.8]{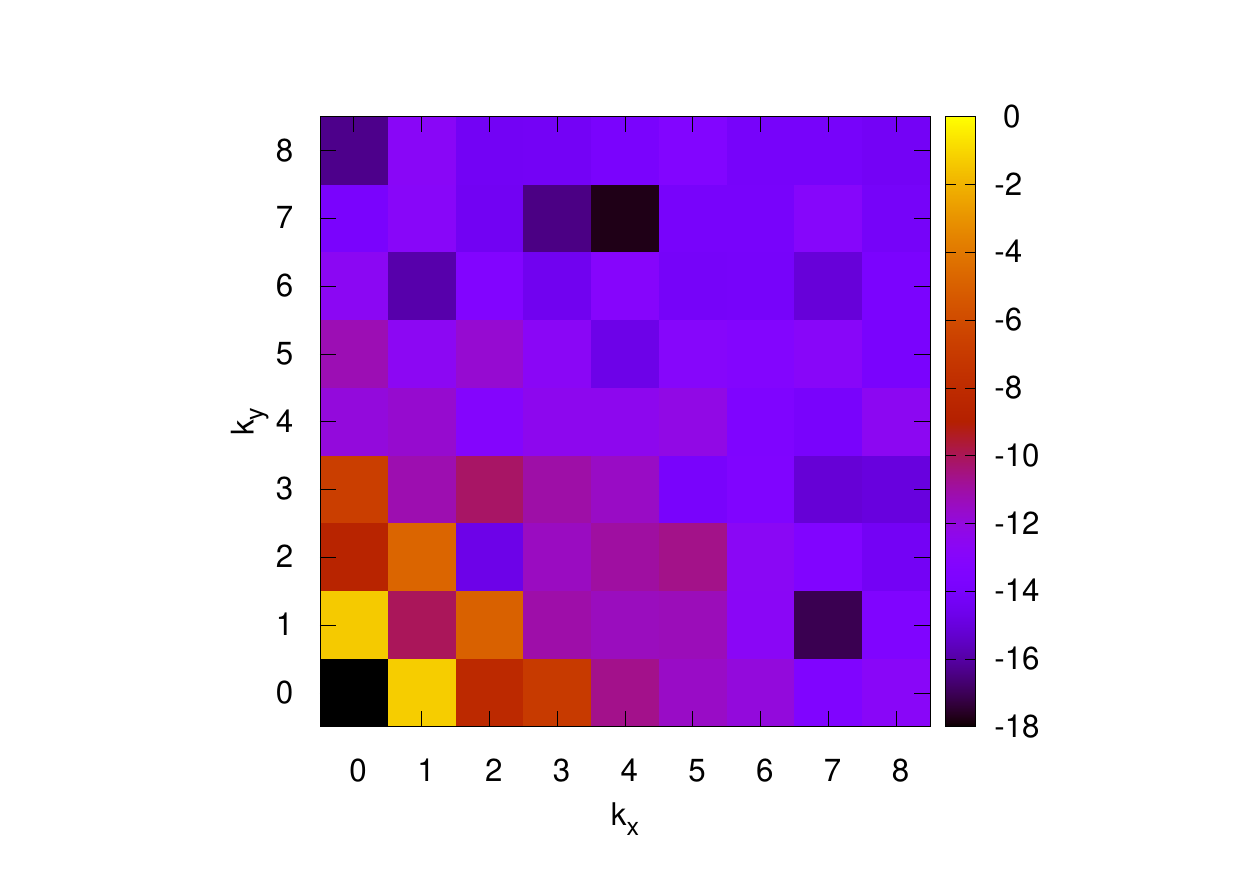}}
\caption{Two-dimensional spectrum of the 
vertical vorticity $\overline{\omega_z}(x,y)$ 
averaged over the homogeneous $z$ direction and time, 
for the run at $F=0.16$. 
The range of wavenumbers is limited to 
$0 \le k_i \le 8$.  
Colors are log-scaled.}
\label{fig6}
\end{figure}

When the vorticity field is averaged also over time it becomes almost monochromatic, 
with a spatial structure equal to that of the laminar solution, i.e.
$\overline{\omega_z}(x,y) \simeq U K (\cos(Kx)+\cos(Ky))$ (see Fig.~\ref{fig5}). 
This remarkable feature is observed also in other turbulent flows forced
by monochromatic forcing, such as the Kolmogorov flow 
\cite{musacchio2014turbulent}.
The 2D spectral analysis of $\overline{\omega_z}(x,y)$, 
reported in Figure~\ref{fig6}, shows that the amplitude 
of the first subleading modes 
$(k_x,k_y)=(1,2)$ and $(k_x,k_y)=(2,1)$ is about 
$15 \%$ of that of the leading modes 
$(k_x,k_y)=(0,1)$ and $(k_x,k_y)=(1,0)$.  
The modes with $k > 3$ are essentially negligible. 

We now consider the vorticity budget.
In statistically stationary
conditions, by taking the vertical average of the $z$-component
of Eq.~(\ref{eq1}), one obtains 
\begin{equation}
\partial_x \overline{u_x \omega_z} + 
\partial_y \overline{u_y \omega_z} -
\partial_x \overline{\omega_x u_z} -
\partial_y \overline{\omega_y u_z} - 
\nu (\partial_x^2 + \partial_y^2) \overline{\omega_z} =
K F (\cos(Kx)+\cos(Ky))
\label{eq2}
\end{equation}
and similar equations for the other two components (without the forcing
term which has the $z$ component only).

In the laminar regime, the first two terms on the 
right-hand side of the equation (advection terms) sum to zero, 
the vortex stretching terms vanish identically, and
Eq.~(\ref{eq2}) reduces to the balance between the forcing and the viscous term,
which gives the relation for the amplitudes $F=\nu K^2 U_0$.
In analogy with \cite{musacchio2014turbulent},
we define the friction factor $f \equiv F/(K U^2)$,
which represents the inverse of the efficiency
of the work done by the forcing to produce the coherent vortical motion.
At $Re < Re_c$, the friction factor
follows the viscous scaling $f =\nu K/U_0 = 1/Re$.

\begin{figure}[h!]
\centerline{\includegraphics[scale=0.8]{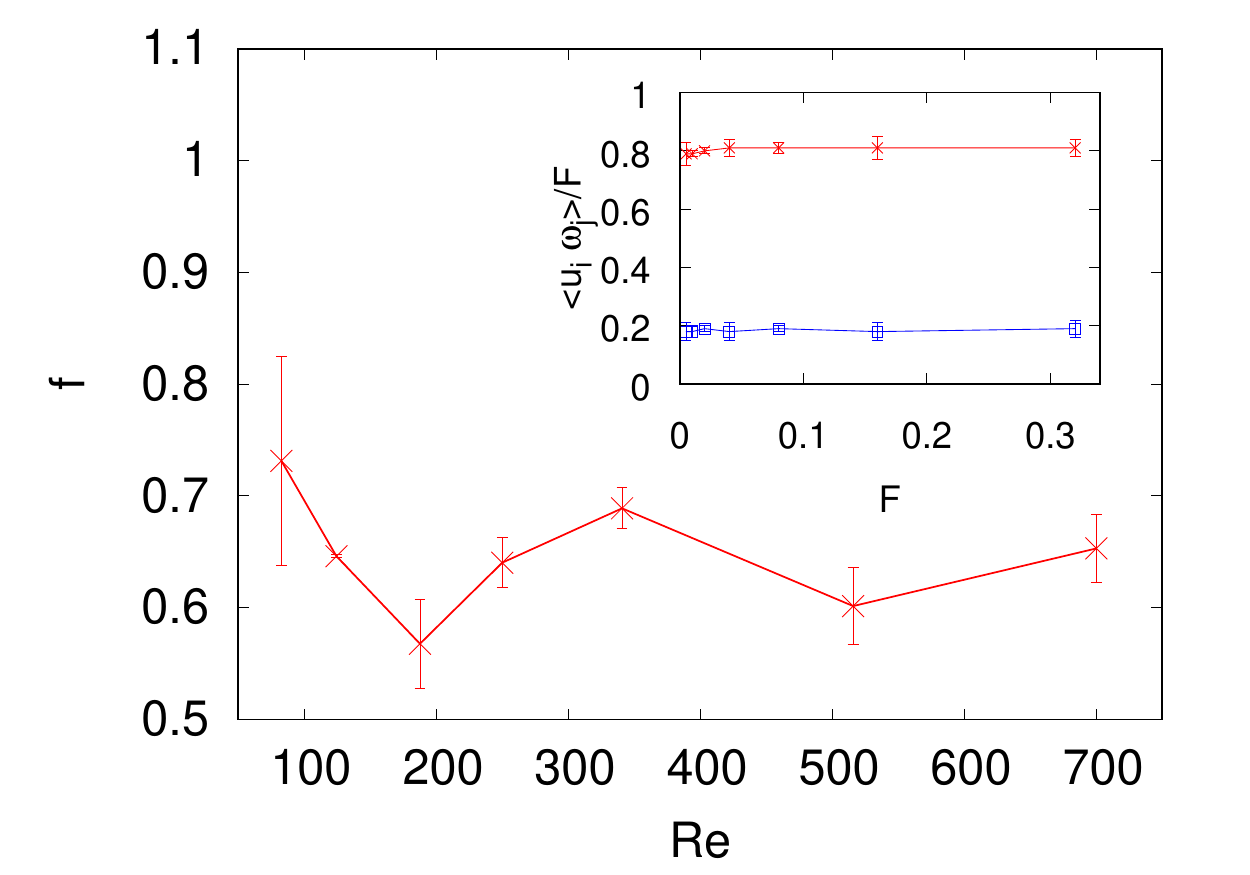}}
\caption{Friction factor $f=F/(K U^2)$ as a function of the Reynolds number
$Re=\nu K/U$. Errors are estimated from the fluctuations of the amplitude
of the average velocity profile.
Inset: relative contribution of the quadratic terms to the momentum
budget of Eq.~(\ref{eq2}), normalized by the forcing amplitude. 
The red upper line corresponds to the contribution 
from the advection terms $\overline{u_i \omega_z}$, the blue lower one
to that from the vortex stretching terms $\overline{\omega_i u_z}$.
}
\label{fig7}
\end{figure}

In the turbulent regime ($Re \gg Re_c$) the contribution
of the dissipative term in Eq.~(\ref{eq2}) becomes negligible
and the forcing is balanced by the quadratic terms.
These terms have a complex structure,
given by the superposition of several modes
beyond the basic mode of the forcing.
This is an important difference with respect to the case
of the Kolmogorov parallel flow where the Reynolds stress has a simple
monochromatic profile \cite{musacchio2014turbulent}.
Nonetheless, given that the cellular forcing is 
monochromatic, it is interesting to consider the contributions
of the leading Fourier components, associated with the wavenumbers 
$(k_x,k_y)=(0,1)$ and $(k_x,k_y)=(1,0)$ where the forcing acts, 
of the different terms to the vorticity budget.
The inset of Fig.~\ref{fig7} shows the relative contribution
to the vorticity budget of the quadratic terms in Eq.~(\ref{eq2}).
These correspond to the sum of the amplitudes of the modes 
$(k_x,k_y)=(0,1)$ and $(k_x,k_y)=(1,0)$ from either the advection terms
$\overline{u_i \omega_z}$ or the vortex-stretching ones 
$\overline{\omega_i u_z}$ (with $i=x,y$), 
normalized by the amplitude of the forcing $F$. 
Independently of $F$, we find that the main contribution (about $80 \%$)
to the vorticity budget, Eq.~(\ref{eq2}), comes from the advection term
${\bm u} \cdot {\bm \nabla \omega}$, while the vortex stretching
term ${\bm \omega} \cdot {\bm \nabla u}$ 
only accounts for the remaining $20 \%$.

The friction factor is shown in the main plot of Fig.~\ref{fig7} 
as a function of the Reynolds number.
The error bars are here estimated from the temporal fluctuations of the 
amplitude $U$ of the modes $(k_x,k_y)=(0,1)$ and $(k_x,k_y)=(1,0)$
of the velocity field. 
These fluctuations correspond to the alternation of configurations
with strong and weak mean flow (see Fig.~\ref{fig5}).
Although the statistical uncertainty on the values of $f$ is quite large
(in particular for the simulation with $F=0.005$),
we observe that at high $Re$ the friction factor fluctuates
around an almost constant value $f \approx 0.65 \pm 0.05$.
This is in agreement with the dimensional scaling $U \sim E^{1/2} \sim F$,
which is suggested by the results of Fig.~\ref{fig2}.
We note that the asymptotic value $f \approx 0.65$
is much larger than that of the Kolmogorov flow, where $f \approx 0.12$,
meaning that the turbulent cellular flow has a larger bulk resistance. 

\subsection{Temporal fluctuations}
\label{sec:fluct}

\begin{figure}[h!]
\centerline{\includegraphics[scale=0.8]{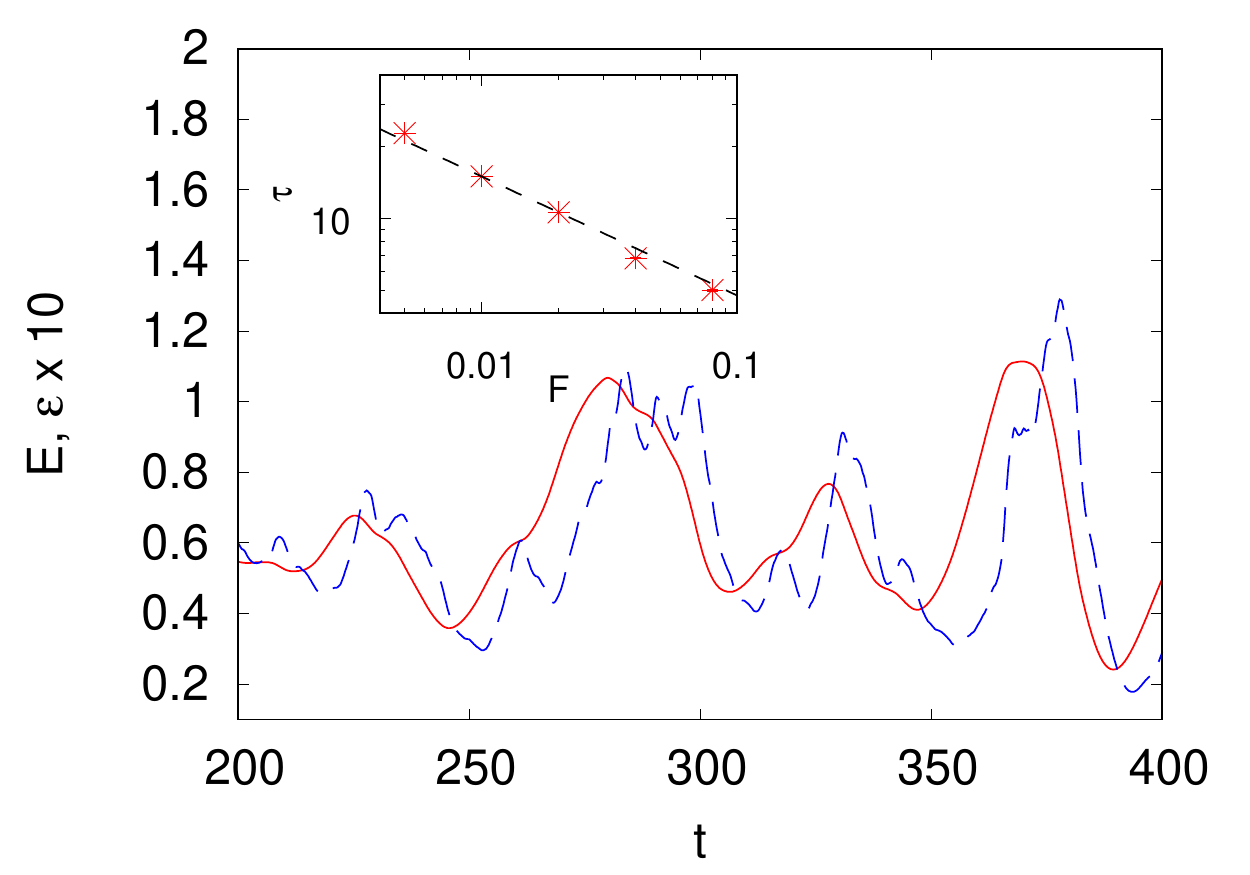}}
\caption{Kinetic energy $E$ (red continuous line) and 
energy dissipation $\varepsilon$ (blue dashed line)
as a function of time for the simulation with $F=0.08$.
For clarity, $\varepsilon$ has been rescaled by a factor $10$.
Inset: correlation time between energy and dissipation (defined 
from the maximum of the correlation function, see text) as a function of $F$. 
The dashed line represents the dimensional scaling $F^{-1/2}$.}
\label{fig8}
\end{figure}

Although the turbulent flow is statistically stationary, 
it displays strong fluctuations in time, 
which are evident by comparing the fields $\overline{\omega_z}(x,y)$ 
at different times (Fig.~\ref{fig5}), 
as well as by looking at the instantaneous values of the 
kinetic energy (Fig.~\ref{fig2}). 
Together with the kinetic energy, also the energy dissipation rate 
$\varepsilon$ is affected by 
important fluctuations. This is shown in Fig.~\ref{fig8}, which 
reports a subset of the full time series
of $E(t)$ and $\varepsilon(t)$ for the simulation
at $F=0.08$. As expected, 
the evolution of $\varepsilon(t)$ is correlated to that of $E(t)$, 
with a time delay 
corresponding to the time for
a fluctuation produced at large scale by the forcing to reach the smallest
scale where it is dissipated by viscosity. 
We measure this time 
lag from the maximum of the 
correlation function between energy and dissipation, defined as 
$C(\tau)=\int E(t) \varepsilon(t+\tau) dt$ (defined for $\tau>0$ since 
$\varepsilon(t)$ is retarded with respect to $E(t)$).
The inset of Fig.~\ref{fig8} shows that the delay time $\tau$ 
computed in our simulations 
scales as the integral time scale $T=E/\varepsilon \simeq F^{-1/2}$. 
The presence of quasi-periodic oscillations of $E(t)$ and  $\varepsilon(t)$, 
with a $\pi/2$ phase delay had been previously reported~\cite{goto2017hierarchy}. 

The strong fluctuations of the global quantities (such as $E$ and $\varepsilon$) 
observed in the turbulent cellular flow have an effect also on the energy spectra. 
Theoretical studies of unsteady turbulence 
performed with two-scale direct-interaction approximation 
method~\cite{yoshizawa1994nonequilibrium} 
and with multiple-scale perturbation method~\cite{woodruff2006multiple} 
predicted that the temporal variation of the energy dissipation rate $\varepsilon(t)$
produces a correction to the Kolmogorov energy spectrum of the form 
$E(k) = E_0(k) + E_1(k)$, 
where $E_0(k) = C \varepsilon^{2/3} k^{-5/3}$  is the Kolmogorov spectrum 
and $E_1(k)$ is the spectral correction given by 
{\begin{equation} 
E_1(k) = C_1 \varepsilon^{-2/3} \dot{\varepsilon} k^{-7/3} \, ,
\label{eq3}
\end{equation} 
where $\dot{\varepsilon} = d \varepsilon / dt$ and $C_1$ is a constant. 

This result can be explained by simple heuristic arguments, 
based on the multiple-scale method~\cite{woodruff2006multiple}. 
The time evolution of the energy spectrum is given by~\cite{batchelor1953theory} 
\begin{equation} 
\partial_t E(k,t) = P(k,t) - T(k,t) - D(k,t) \, ,
\label{eq4} 
\end{equation}
where $P(k,t)$ is the production spectrum due to the external force, 
$T(k,t)$ is the energy transfer 
and $D(k,t) = 2 \nu k^2 E(k,t)$ is the viscous dissipation spectrum. 
A dimensional estimate for the energy transfer gives 
$T(k,t) \sim E(k,t)/\tau_k$, where 
$\tau_k  =\left[ \int_0^k q^2 E(q,t) dq\right]^{-1/2}$ is the 
characteristic distortion time of the eddies of size $1/k$.
The hypothesis of a steady state with constant energy flux 
$\mathcal{F}(k) = \int_0^k T(q,t) dq = \varepsilon$ in the inertial range, 
equal to the energy dissipation rate, leads to the dimensional relation
$\mathcal{F}(k)  \sim k T(k,t) \sim k E(k,t) [ k^3 E(k,t) ]^{1/2} \sim \varepsilon$, 
which gives the Kolmogorov spectrum 
$E(k) = C \varepsilon^{2/3} k^{-5/3}$. 

Now, let us assume that the energy dissipation rate $\varepsilon$
varies slowly in time, with a characteristic time of order $1/\epsilon$ 
(with $\epsilon$ arbitrarily small). 
Following the multiple-scale approach, 
we introduce the {\it slow time} variable $t' = \epsilon t$, 
such that the time derivative becomes $\partial_t = \epsilon \partial_{t'}$, 
and we expand the energy spectrum as 
$E(k,t') = E_0(k,t') + \epsilon E_1(k,t')$
where the zeroth-order approximation 
$E_0(k,t') = C \varepsilon(t')^{2/3} k^{-5/3}$ 
is the Kolmogorov solution, 
and $E_1(k,t')$ is the first order correction. 
Inserting this expansion in Eq.~(\ref{eq4})
and assuming that in the inertial range the energy dissipation and production can be neglected, 
we get at first order in $\epsilon$ the equation:
\begin{equation}
\partial_{t'} E_0(k,t') = E_1(k,t')  \left[ \int_0^k q^2 E_0(q,t') dq\right]^{1/2}  
+ \frac{1}{2} E_0(k,t') \left[ \int_0^k q^2 E_1(q,t') dq\right] \left[ \int_0^k q^2 E_0(q,t') dq\right]^{-1/2} \, .
\label{eq5}
\end{equation}
Using the explicit 
expression of $E_0$ and assuming a power-law form for the 
correction $E_1(k) \sim k^{\beta}$, we obtain the dimensional relation
$\varepsilon^{-1/3} \dot{\varepsilon} k^{-5/3} \sim 
E_1 \varepsilon^{1/3} k^{2/3}$,  
which leads to the prediction
$E_1(k) \sim  \varepsilon^{-2/3} \dot{\varepsilon} k^{-7/3}$
for the first-order spectral correction. 
We note that this dimensional argument is independent of the specific choice
of the closure for the energy transfer in terms of the energy spectrum. The
same prediction can be obtained
by using different 
closures~\cite{rubinstein2005self,woodruff2006multiple,bos2017dissipation,steiros2022turbulence}. 
We also note that, for the sake of simplicity, here we assumed statistical
isotropy and homogeneity of the flow.  A refined theory which accounts for the
presence of spatial inhomogeneities has been recently
proposed~\cite{bos2020production,araki2022inertial}. 

\begin{figure}[h!]
\centerline{\includegraphics[scale=0.8]{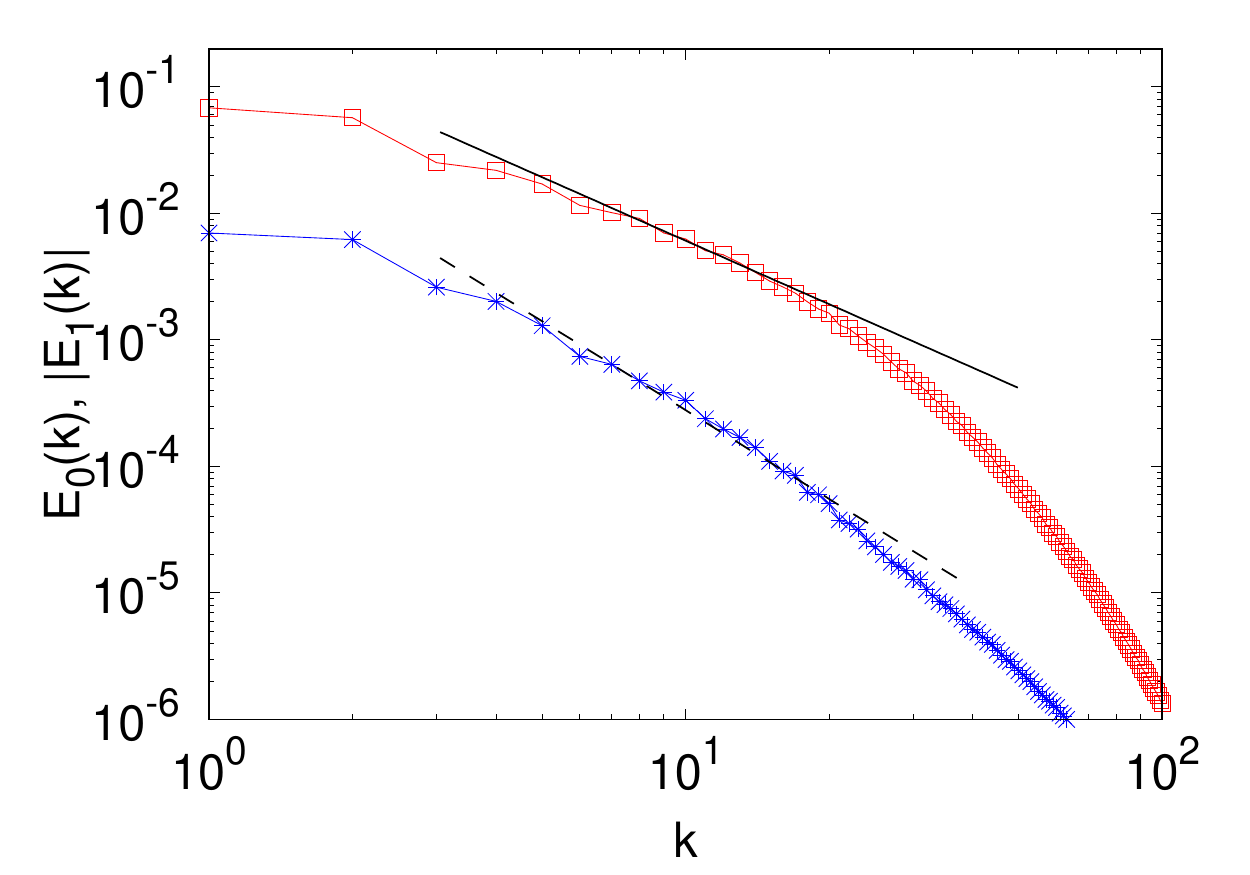}}
\caption{
Isotropic energy spectrum $E_0(k)$ (red squares) and 
spectral correction $E_1(k)$, in absolute value (blue crosses). 
The continuous line is the Kolmogorov spectrum
$E_0(k)=C \varepsilon^{2/3} k^{-5/3}$ with $C=2.0$,
the dashed line represents the $k^{-7/3}$ scaling.
}
\label{fig9}
\end{figure}
 
Note that the sign of the correction $E_1(k)$ depends on the time derivative of $\varepsilon$. 
This has a simple physical interpretation.
As we see in Fig.~\ref{fig8}, the maxima of the energy occur at times for which $\dot{\varepsilon} > 0$. 
Indeed, from the energy balance we have that $\dot{\varepsilon} \simeq - \ddot{E}$ 
(neglecting the time variation of the energy input),  
and thus when $\dot{\varepsilon}>0$ the forcing accumulates energy 
in the large-scale modes, producing a steepening of the energy spectrum
which is in agreement with a positive sign of the correction $E_1$. 
Vice versa, the decrease of the energy dissipation rate ($\dot{\varepsilon} < 0$) 
occurs immediately after strong dissipation events, 
during which a large amount of energy is removed from the large scales and 
it is transferred to the viscous scales. 
In this phase, the negative sign of the correction $E_1$ 
results in a flatter energy spectrum. 

This observation suggests an efficient method~\cite{horiuti2013nonequilibrium} 
to extract the subleading term $E_1(k,t)$ in our numerical simulations. 
Computing the conditional spectra $E^+(k) = E_0(k) + |E_1(k)| $ 
averaged during the time intervals in which $\dot{\varepsilon} > 0$ 
and $E^-(k) = E_0(k) - |E_1(k)| $ when $\dot{\varepsilon} < 0$, 
the correction can be easily obtained as $|E_1(k)| = \frac{1}{2} [E^+(k)-E^-(k)]$. 
The result is shown in Fig.~\ref{fig9}, which displays a clear $k^{-7/3}$ scaling. 

\section{Conclusions}
\label{sec:concl}

We presented a series of direct numerical simulations of a turbulent, 
inhomogeneous flow, sustained by a two-dimensional, two-component force 
with a monochromatic, cellular (vortex-like) spatial structure.  
Notwithstanding the large-scale anisotropy imposed by the forcing, in the
statistically steady regime at high $Re$, the velocity field is found to
display almost complete isotropy at small scales, which was assessed by
analyzing 1D energy spectra.

Our results show, however, that
the turbulent flow keeps memory of the spatial structure
of the forcing, which is recovered by averaging the vorticity field 
along the vertical homogeneous direction 
and/or over time. 
As a consequence, the mean flow has basically the same form of 
the laminar solution, except for a reduced amplitude.  This
property allows us to define a friction factor, in analogy with the case of
wall-bounded flows, which quantifies the intensity of the force that is
necessary to sustain a mean flow with given kinetic energy. 
Our results show that the friction factor approaches a constant value 
at large $Re$, consistently with dimensional predictions.

Besides the time-averaged statistical properties, 
we find that the turbulent cellular flow displays strong temporal fluctuations.
Its dynamics are characterized by an alternation of phases in which the 2D
forcing accumulates kinetic energy in the large-scale flow,
followed by phases in which the energy is rapidly transferred
toward the small dissipative scales by the 3D turbulent cascade process.
In the ``energy charging'' phase, the flow displays a marked 2D structure,
which is the hallmark of the forcing.   
Conversely, the 2D mean flow is weaker in the ``discharging'' phases,
which are characterized by intense energy dissipation rates.
The temporal fluctuations of the dissipation rates affect also
the spectral properties, causing the appearance of a correction
to the Kolmogorov energy spectrum,
with subleading spectral exponent $-7/3$,
in agreement with theoretical predictions
\cite{yoshizawa1994nonequilibrium,woodruff2006multiple}.

We conclude by noting that the 
turbulent cellular flow can offer an ideal
configuration to study the spreading of tracer or
inertial particles in an inhomogenous turbulent flow for,
e.g., applications to pollutant dispersion and 
plastics sedimentation~\cite{baudena2022streaming}, one of the 
big challenges in environmental fluid dynamics 
\cite{dauxois2021confronting}, avoiding
ad-hoc small-scale parametrizations. 
It can be also instructive for other Lagrangian advection studies, which are
often considered to assess the effect of unresolved velocity
components, but rely on kinematic or essentially
laminar flows~\cite{forgia2022numerical}.  Similarly, it 
may reveal interesting for fundamental studies about the role of
turbulence on plankton dynamics, particularly in relation to nutrient
upwelling~\cite{GF2020,FFM2022} and to the different assumptions of the
biological modeling. 
Along a similar line of reasoning, it can be noted that 
cellular-flow patterns (with vortical motions on the vertical) 
also appear in Langmuir
circulation~\cite{CL1976,thorpe2004}, a distinctive feature of the ocean
mixed-layer dynamics with important implications on marine ecology.  Starting
from the basic fluid-dynamical equations, a reduced, kinematic model of such
motions was otained~\cite{CL1976,ML1985,bees1998}, which essentially
coincides with the laminar solution of the flow considered here.
Therefore, the present setup may also allow to explore how 
the survival conditions of
phytoplankton are affected by the simultaneous presence of large-scale
advection and small-scale turbulent motions in (Langmuir) convective
flows~\cite{TF2011,LVM2017,TCMB2021}.

\section*{Acknowledgments}
We acknowledge HPC CINECA for computing resources
(INFN-CINECA grant no. INFN22-FieldTurb).
SB acknowledges support from INFN during a scientific stay at 
the Deparment of Physics, University of Torino.

\bibliography{biblio}

\begin{thebibliography}{32}%
\makeatletter
\providecommand \@ifxundefined [1]{%
 \@ifx{#1\undefined}
}%
\providecommand \@ifnum [1]{%
 \ifnum #1\expandafter \@firstoftwo
 \else \expandafter \@secondoftwo
 \fi
}%
\providecommand \@ifx [1]{%
 \ifx #1\expandafter \@firstoftwo
 \else \expandafter \@secondoftwo
 \fi
}%
\providecommand \natexlab [1]{#1}%
\providecommand \enquote  [1]{``#1''}%
\providecommand \bibnamefont  [1]{#1}%
\providecommand \bibfnamefont [1]{#1}%
\providecommand \citenamefont [1]{#1}%
\providecommand \href@noop [0]{\@secondoftwo}%
\providecommand \href [0]{\begingroup \@sanitize@url \@href}%
\providecommand \@href[1]{\@@startlink{#1}\@@href}%
\providecommand \@@href[1]{\endgroup#1\@@endlink}%
\providecommand \@sanitize@url [0]{\catcode `\\12\catcode `\$12\catcode
  `\&12\catcode `\#12\catcode `\^12\catcode `\_12\catcode `\%12\relax}%
\providecommand \@@startlink[1]{}%
\providecommand \@@endlink[0]{}%
\providecommand \url  [0]{\begingroup\@sanitize@url \@url }%
\providecommand \@url [1]{\endgroup\@href {#1}{\urlprefix }}%
\providecommand \urlprefix  [0]{URL }%
\providecommand \Eprint [0]{\href }%
\providecommand \doibase [0]{http://dx.doi.org/}%
\providecommand \selectlanguage [0]{\@gobble}%
\providecommand \bibinfo  [0]{\@secondoftwo}%
\providecommand \bibfield  [0]{\@secondoftwo}%
\providecommand \translation [1]{[#1]}%
\providecommand \BibitemOpen [0]{}%
\providecommand \bibitemStop [0]{}%
\providecommand \bibitemNoStop [0]{.\EOS\space}%
\providecommand \EOS [0]{\spacefactor3000\relax}%
\providecommand \BibitemShut  [1]{\csname bibitem#1\endcsname}%
\let\auto@bib@innerbib\@empty
\bibitem [{\citenamefont {Avila}\ \emph {et~al.}(2022)\citenamefont {Avila},
  \citenamefont {Barkley},\ and\ \citenamefont {Hof}}]{avila2022transition}%
  \BibitemOpen
  \bibfield  {author} {\bibinfo {author} {\bibfnamefont {Marc}\ \bibnamefont
  {Avila}}, \bibinfo {author} {\bibfnamefont {Dwight}\ \bibnamefont {Barkley}},
  \ and\ \bibinfo {author} {\bibfnamefont {Bj{\"o}rn}\ \bibnamefont {Hof}},\
  }\bibfield  {title} {\enquote {\bibinfo {title} {Transition to turbulence in
  pipe flow},}\ }\href@noop {} {\bibfield  {journal} {\bibinfo  {journal}
  {Annu. Rev. Fluid Mech.}\ }\textbf {\bibinfo {volume} {55}},\ \bibinfo
  {pages} {192} (\bibinfo {year} {2022})}\BibitemShut {NoStop}%
\bibitem [{\citenamefont {Grossmann}\ \emph {et~al.}(2016)\citenamefont
  {Grossmann}, \citenamefont {Lohse},\ and\ \citenamefont
  {Sun}}]{grossmann2016high}%
  \BibitemOpen
  \bibfield  {author} {\bibinfo {author} {\bibfnamefont {Siegfried}\
  \bibnamefont {Grossmann}}, \bibinfo {author} {\bibfnamefont {Detlef}\
  \bibnamefont {Lohse}}, \ and\ \bibinfo {author} {\bibfnamefont {Chao}\
  \bibnamefont {Sun}},\ }\bibfield  {title} {\enquote {\bibinfo {title}
  {High--reynolds number taylor-couette turbulence},}\ }\href@noop {}
  {\bibfield  {journal} {\bibinfo  {journal} {Annu. Rev. Fluid Mech.}\ }\textbf
  {\bibinfo {volume} {48}},\ \bibinfo {pages} {53--80} (\bibinfo {year}
  {2016})}\BibitemShut {NoStop}%
\bibitem [{\citenamefont {Meshalkin}\ and\ \citenamefont
  {Sinai}(1961)}]{meshalkin1961investigation}%
  \BibitemOpen
  \bibfield  {author} {\bibinfo {author} {\bibfnamefont {LD}~\bibnamefont
  {Meshalkin}}\ and\ \bibinfo {author} {\bibfnamefont {IG}~\bibnamefont
  {Sinai}},\ }\bibfield  {title} {\enquote {\bibinfo {title} {{Investigation of
  the stability of a stationary solution of a system of equations for the plane
  movement of an incompressible viscous liquid}},}\ }\href@noop {} {\bibfield
  {journal} {\bibinfo  {journal} {J. Applied Math. Mech.}\ }\textbf {\bibinfo
  {volume} {25}},\ \bibinfo {pages} {1700--1705} (\bibinfo {year}
  {1961})}\BibitemShut {NoStop}%
\bibitem [{\citenamefont {Musacchio}\ and\ \citenamefont
  {Boffetta}(2014)}]{musacchio2014turbulent}%
  \BibitemOpen
  \bibfield  {author} {\bibinfo {author} {\bibfnamefont {S}~\bibnamefont
  {Musacchio}}\ and\ \bibinfo {author} {\bibfnamefont {G}~\bibnamefont
  {Boffetta}},\ }\bibfield  {title} {\enquote {\bibinfo {title} {{Turbulent
  channel without boundaries: The periodic Kolmogorov flow}},}\ }\href@noop {}
  {\bibfield  {journal} {\bibinfo  {journal} {Phys. Rev. E}\ }\textbf {\bibinfo
  {volume} {89}},\ \bibinfo {pages} {023004} (\bibinfo {year}
  {2014})}\BibitemShut {NoStop}%
\bibitem [{\citenamefont {Taylor}\ and\ \citenamefont
  {Green}(1937)}]{taylor1937mechanism}%
  \BibitemOpen
  \bibfield  {author} {\bibinfo {author} {\bibfnamefont {G~I}\ \bibnamefont
  {Taylor}}\ and\ \bibinfo {author} {\bibfnamefont {A~E}\ \bibnamefont
  {Green}},\ }\bibfield  {title} {\enquote {\bibinfo {title} {Mechanism of the
  production of small eddies from large ones},}\ }\href@noop {} {\bibfield
  {journal} {\bibinfo  {journal} {Proc. R. Soc. London A}\ }\textbf {\bibinfo
  {volume} {158}},\ \bibinfo {pages} {499--521} (\bibinfo {year}
  {1937})}\BibitemShut {NoStop}%
\bibitem [{\citenamefont {Brachet}\ \emph {et~al.}(1983)\citenamefont
  {Brachet}, \citenamefont {Meiron}, \citenamefont {Orszag}, \citenamefont
  {Nickel}, \citenamefont {Morf},\ and\ \citenamefont
  {Frisch}}]{brachet1983small}%
  \BibitemOpen
  \bibfield  {author} {\bibinfo {author} {\bibfnamefont {Marc~E}\ \bibnamefont
  {Brachet}}, \bibinfo {author} {\bibfnamefont {Daniel~I}\ \bibnamefont
  {Meiron}}, \bibinfo {author} {\bibfnamefont {Steven~A}\ \bibnamefont
  {Orszag}}, \bibinfo {author} {\bibfnamefont {BG}~\bibnamefont {Nickel}},
  \bibinfo {author} {\bibfnamefont {Rudolf~H}\ \bibnamefont {Morf}}, \ and\
  \bibinfo {author} {\bibfnamefont {Uriel}\ \bibnamefont {Frisch}},\ }\bibfield
   {title} {\enquote {\bibinfo {title} {{Small-scale structure of the
  Taylor--Green vortex}},}\ }\href@noop {} {\bibfield  {journal} {\bibinfo
  {journal} {J. Fluid Mech.}\ }\textbf {\bibinfo {volume} {130}},\ \bibinfo
  {pages} {411--452} (\bibinfo {year} {1983})}\BibitemShut {NoStop}%
\bibitem [{\citenamefont {Arnold}(1965)}]{arnold1965topologie}%
  \BibitemOpen
  \bibfield  {author} {\bibinfo {author} {\bibfnamefont {Vladimir~I}\
  \bibnamefont {Arnold}},\ }\bibfield  {title} {\enquote {\bibinfo {title} {Sur
  la topologie des {\'e}coulements stationnaires des fluides parfaits},}\ }in\
  \href@noop {} {\emph {\bibinfo {booktitle} {Vladimir I. Arnold-Collected
  Works}}}\ (\bibinfo  {publisher} {Springer},\ \bibinfo {year} {1965})\ pp.\
  \bibinfo {pages} {15--18}\BibitemShut {NoStop}%
\bibitem [{\citenamefont {Goto}\ \emph {et~al.}(2017)\citenamefont {Goto},
  \citenamefont {Saito},\ and\ \citenamefont {Kawahara}}]{goto2017hierarchy}%
  \BibitemOpen
  \bibfield  {author} {\bibinfo {author} {\bibfnamefont {Susumu}\ \bibnamefont
  {Goto}}, \bibinfo {author} {\bibfnamefont {Yuta}\ \bibnamefont {Saito}}, \
  and\ \bibinfo {author} {\bibfnamefont {Genta}\ \bibnamefont {Kawahara}},\
  }\bibfield  {title} {\enquote {\bibinfo {title} {Hierarchy of antiparallel
  vortex tubes in spatially periodic turbulence at high reynolds numbers},}\
  }\href@noop {} {\bibfield  {journal} {\bibinfo  {journal} {Phys. Rev.
  Fluids}\ }\textbf {\bibinfo {volume} {2}},\ \bibinfo {pages} {064603}
  (\bibinfo {year} {2017})}\BibitemShut {NoStop}%
\bibitem [{\citenamefont {Majda}\ and\ \citenamefont
  {Wang}(2006)}]{majda2006nonlinear}%
  \BibitemOpen
  \bibfield  {author} {\bibinfo {author} {\bibfnamefont {Andrew}\ \bibnamefont
  {Majda}}\ and\ \bibinfo {author} {\bibfnamefont {Xiaoming}\ \bibnamefont
  {Wang}},\ }\href@noop {} {\emph {\bibinfo {title} {Nonlinear dynamics and
  statistical theories for basic geophysical flows}}}\ (\bibinfo  {publisher}
  {Cambridge University Press},\ \bibinfo {year} {2006})\BibitemShut {NoStop}%
\bibitem [{\citenamefont {{La Forgia}}\ \emph {et~al.}(2022)\citenamefont {{La
  Forgia}}, \citenamefont {Cavaliere}, \citenamefont {Espa}, \citenamefont
  {Falcini},\ and\ \citenamefont {Lacorata}}]{forgia2022numerical}%
  \BibitemOpen
  \bibfield  {author} {\bibinfo {author} {\bibfnamefont {G}~\bibnamefont {{La
  Forgia}}}, \bibinfo {author} {\bibfnamefont {D}~\bibnamefont {Cavaliere}},
  \bibinfo {author} {\bibfnamefont {S}~\bibnamefont {Espa}}, \bibinfo {author}
  {\bibfnamefont {F}~\bibnamefont {Falcini}}, \ and\ \bibinfo {author}
  {\bibfnamefont {G}~\bibnamefont {Lacorata}},\ }\bibfield  {title} {\enquote
  {\bibinfo {title} {Numerical and experimental analysis of {Lagrangian}
  dispersion in two-dimensional chaotic flows},}\ }\href@noop {} {\bibfield
  {journal} {\bibinfo  {journal} {Sci. Rep.}\ }\textbf {\bibinfo {volume}
  {12}},\ \bibinfo {pages} {1--12} (\bibinfo {year} {2022})}\BibitemShut
  {NoStop}%
\bibitem [{\citenamefont {Sivashinsky}\ and\ \citenamefont
  {Yakhot}(1985)}]{sivashinsky1985negative}%
  \BibitemOpen
  \bibfield  {author} {\bibinfo {author} {\bibfnamefont {G}~\bibnamefont
  {Sivashinsky}}\ and\ \bibinfo {author} {\bibfnamefont {V}~\bibnamefont
  {Yakhot}},\ }\bibfield  {title} {\enquote {\bibinfo {title} {{Negative
  viscosity effect in large-scale flows}},}\ }\href@noop {} {\bibfield
  {journal} {\bibinfo  {journal} {Phys. Fluids}\ }\textbf {\bibinfo {volume}
  {28}},\ \bibinfo {pages} {1040--1042} (\bibinfo {year} {1985})}\BibitemShut
  {NoStop}%
\bibitem [{\citenamefont {Perlekar}\ and\ \citenamefont
  {Pandit}(2010)}]{perlekar2010turbulence}%
  \BibitemOpen
  \bibfield  {author} {\bibinfo {author} {\bibfnamefont {Prasad}\ \bibnamefont
  {Perlekar}}\ and\ \bibinfo {author} {\bibfnamefont {Rahul}\ \bibnamefont
  {Pandit}},\ }\bibfield  {title} {\enquote {\bibinfo {title}
  {Turbulence-induced melting of a nonequilibrium vortex crystal in a forced
  thin fluid film},}\ }\href@noop {} {\bibfield  {journal} {\bibinfo  {journal}
  {New J. Phys.}\ }\textbf {\bibinfo {volume} {12}},\ \bibinfo {pages} {023033}
  (\bibinfo {year} {2010})}\BibitemShut {NoStop}%
\bibitem [{\citenamefont {Batchelor}(1953)}]{batchelor1953theory}%
  \BibitemOpen
  \bibfield  {author} {\bibinfo {author} {\bibfnamefont {George~Keith}\
  \bibnamefont {Batchelor}},\ }\href@noop {} {\emph {\bibinfo {title} {The
  theory of homogeneous turbulence}}}\ (\bibinfo  {publisher} {Cambridge
  university press},\ \bibinfo {year} {1953})\BibitemShut {NoStop}%
\bibitem [{\citenamefont {Yoshizawa}(1994)}]{yoshizawa1994nonequilibrium}%
  \BibitemOpen
  \bibfield  {author} {\bibinfo {author} {\bibfnamefont {Akira}\ \bibnamefont
  {Yoshizawa}},\ }\bibfield  {title} {\enquote {\bibinfo {title}
  {Nonequilibrium effect of the turbulent-energy-production process on the
  inertial-range energy spectrum},}\ }\href@noop {} {\bibfield  {journal}
  {\bibinfo  {journal} {Phys. Rev. E}\ }\textbf {\bibinfo {volume} {49}},\
  \bibinfo {pages} {4065} (\bibinfo {year} {1994})}\BibitemShut {NoStop}%
\bibitem [{\citenamefont {Woodruff}\ and\ \citenamefont
  {Rubinstein}(2006)}]{woodruff2006multiple}%
  \BibitemOpen
  \bibfield  {author} {\bibinfo {author} {\bibfnamefont {Stephen~L}\
  \bibnamefont {Woodruff}}\ and\ \bibinfo {author} {\bibfnamefont {Robert}\
  \bibnamefont {Rubinstein}},\ }\bibfield  {title} {\enquote {\bibinfo {title}
  {Multiple-scale perturbation analysis of slowly evolving turbulence},}\
  }\href@noop {} {\bibfield  {journal} {\bibinfo  {journal} {J. Fluid Mech.}\
  }\textbf {\bibinfo {volume} {565}},\ \bibinfo {pages} {95--103} (\bibinfo
  {year} {2006})}\BibitemShut {NoStop}%
\bibitem [{\citenamefont {Rubinstein}\ and\ \citenamefont
  {Clark}(2005)}]{rubinstein2005self}%
  \BibitemOpen
  \bibfield  {author} {\bibinfo {author} {\bibfnamefont {Robert}\ \bibnamefont
  {Rubinstein}}\ and\ \bibinfo {author} {\bibfnamefont {Timothy~T}\
  \bibnamefont {Clark}},\ }\bibfield  {title} {\enquote {\bibinfo {title}
  {Self-similar turbulence evolution and the dissipation rate transport
  equation},}\ }\href@noop {} {\bibfield  {journal} {\bibinfo  {journal} {Phys.
  Fluids}\ }\textbf {\bibinfo {volume} {17}},\ \bibinfo {pages} {095104}
  (\bibinfo {year} {2005})}\BibitemShut {NoStop}%
\bibitem [{\citenamefont {Bos}\ and\ \citenamefont
  {Rubinstein}(2017)}]{bos2017dissipation}%
  \BibitemOpen
  \bibfield  {author} {\bibinfo {author} {\bibfnamefont {Wouter~JT}\
  \bibnamefont {Bos}}\ and\ \bibinfo {author} {\bibfnamefont {Robert}\
  \bibnamefont {Rubinstein}},\ }\bibfield  {title} {\enquote {\bibinfo {title}
  {Dissipation in unsteady turbulence},}\ }\href@noop {} {\bibfield  {journal}
  {\bibinfo  {journal} {Phys. Rev. Fluids}\ }\textbf {\bibinfo {volume} {2}},\
  \bibinfo {pages} {022601(R)} (\bibinfo {year} {2017})}\BibitemShut {NoStop}%
\bibitem [{\citenamefont {Steiros}(2022)}]{steiros2022turbulence}%
  \BibitemOpen
  \bibfield  {author} {\bibinfo {author} {\bibfnamefont {K}~\bibnamefont
  {Steiros}},\ }\bibfield  {title} {\enquote {\bibinfo {title} {Turbulence near
  initial conditions},}\ }\href@noop {} {\bibfield  {journal} {\bibinfo
  {journal} {Phys. Rev. Fluids}\ }\textbf {\bibinfo {volume} {7}},\ \bibinfo
  {pages} {104607} (\bibinfo {year} {2022})}\BibitemShut {NoStop}%
\bibitem [{\citenamefont {Bos}(2020)}]{bos2020production}%
  \BibitemOpen
  \bibfield  {author} {\bibinfo {author} {\bibfnamefont {Wouter~JT}\
  \bibnamefont {Bos}},\ }\bibfield  {title} {\enquote {\bibinfo {title}
  {Production and dissipation of kinetic energy in grid turbulence},}\
  }\href@noop {} {\bibfield  {journal} {\bibinfo  {journal} {Phys. Rev.
  Fluids}\ }\textbf {\bibinfo {volume} {5}},\ \bibinfo {pages} {104607}
  (\bibinfo {year} {2020})}\BibitemShut {NoStop}%
\bibitem [{\citenamefont {Araki}\ and\ \citenamefont
  {Bos}(2022)}]{araki2022inertial}%
  \BibitemOpen
  \bibfield  {author} {\bibinfo {author} {\bibfnamefont {Ryo}\ \bibnamefont
  {Araki}}\ and\ \bibinfo {author} {\bibfnamefont {Wouter~JT}\ \bibnamefont
  {Bos}},\ }\bibfield  {title} {\enquote {\bibinfo {title} {Inertial range
  scaling of inhomogeneous turbulence},}\ }\href@noop {} {\bibfield  {journal}
  {\bibinfo  {journal} {arXiv preprint arXiv:2210.14516}\ } (\bibinfo {year}
  {2022})}\BibitemShut {NoStop}%
\bibitem [{\citenamefont {Horiuti}\ and\ \citenamefont
  {Tamaki}(2013)}]{horiuti2013nonequilibrium}%
  \BibitemOpen
  \bibfield  {author} {\bibinfo {author} {\bibfnamefont {Kiyosi}\ \bibnamefont
  {Horiuti}}\ and\ \bibinfo {author} {\bibfnamefont {Takahiro}\ \bibnamefont
  {Tamaki}},\ }\bibfield  {title} {\enquote {\bibinfo {title} {Nonequilibrium
  energy spectrum in the subgrid-scale one-equation model in large-eddy
  simulation},}\ }\href@noop {} {\bibfield  {journal} {\bibinfo  {journal}
  {Phys. Fluids}\ }\textbf {\bibinfo {volume} {25}},\ \bibinfo {pages} {125104}
  (\bibinfo {year} {2013})}\BibitemShut {NoStop}%
\bibitem [{\citenamefont {Baudena}\ \emph {et~al.}(2022)\citenamefont
  {Baudena}, \citenamefont {Ser-Giacomi}, \citenamefont {Jal{\'o}n-Rojas},
  \citenamefont {Galgani},\ and\ \citenamefont
  {Pedrotti}}]{baudena2022streaming}%
  \BibitemOpen
  \bibfield  {author} {\bibinfo {author} {\bibfnamefont {A}~\bibnamefont
  {Baudena}}, \bibinfo {author} {\bibfnamefont {E}~\bibnamefont {Ser-Giacomi}},
  \bibinfo {author} {\bibfnamefont {I}~\bibnamefont {Jal{\'o}n-Rojas}},
  \bibinfo {author} {\bibfnamefont {F}~\bibnamefont {Galgani}}, \ and\ \bibinfo
  {author} {\bibfnamefont {M~L}\ \bibnamefont {Pedrotti}},\ }\bibfield  {title}
  {\enquote {\bibinfo {title} {The streaming of plastic in the {Mediterranean
  Sea}},}\ }\href@noop {} {\bibfield  {journal} {\bibinfo  {journal} {Nat.
  Commun.}\ }\textbf {\bibinfo {volume} {13}},\ \bibinfo {pages} {1--9}
  (\bibinfo {year} {2022})}\BibitemShut {NoStop}%
\bibitem [{\citenamefont {Dauxois}\ \emph {et~al.}(2021)\citenamefont
  {Dauxois}, \citenamefont {Peacock}, \citenamefont {Bauer}, \citenamefont
  {Caulfield}, \citenamefont {Cenedese}, \citenamefont {Gorle}, \citenamefont
  {Haller}, \citenamefont {Ivey}, \citenamefont {Linden}, \citenamefont
  {Meiburg}, \citenamefont {Pinardi}, \citenamefont {Vriend},\ and\
  \citenamefont {Woods}}]{dauxois2021confronting}%
  \BibitemOpen
  \bibfield  {author} {\bibinfo {author} {\bibfnamefont {T.}~\bibnamefont
  {Dauxois}}, \bibinfo {author} {\bibfnamefont {T.}~\bibnamefont {Peacock}},
  \bibinfo {author} {\bibfnamefont {P.}~\bibnamefont {Bauer}}, \bibinfo
  {author} {\bibfnamefont {C.P.}\ \bibnamefont {Caulfield}}, \bibinfo {author}
  {\bibfnamefont {C.}~\bibnamefont {Cenedese}}, \bibinfo {author}
  {\bibfnamefont {C.}~\bibnamefont {Gorle}}, \bibinfo {author} {\bibfnamefont
  {G.}~\bibnamefont {Haller}}, \bibinfo {author} {\bibfnamefont {G.N.}\
  \bibnamefont {Ivey}}, \bibinfo {author} {\bibfnamefont {P.F.}\ \bibnamefont
  {Linden}}, \bibinfo {author} {\bibfnamefont {E.}~\bibnamefont {Meiburg}},
  \bibinfo {author} {\bibfnamefont {N.}~\bibnamefont {Pinardi}}, \bibinfo
  {author} {\bibfnamefont {N.M.}\ \bibnamefont {Vriend}}, \ and\ \bibinfo
  {author} {\bibfnamefont {A.W.}\ \bibnamefont {Woods}},\ }\bibfield  {title}
  {\enquote {\bibinfo {title} {Confronting grand challenges in environmental
  fluid mechanics},}\ }\href@noop {} {\bibfield  {journal} {\bibinfo  {journal}
  {Phys. Rev. Fluids}\ }\textbf {\bibinfo {volume} {6}},\ \bibinfo {pages}
  {020501} (\bibinfo {year} {2021})}\BibitemShut {NoStop}%
\bibitem [{\citenamefont {Guseva}\ and\ \citenamefont {Feudel}(2020)}]{GF2020}%
  \BibitemOpen
  \bibfield  {author} {\bibinfo {author} {\bibfnamefont {K}~\bibnamefont
  {Guseva}}\ and\ \bibinfo {author} {\bibfnamefont {U}~\bibnamefont {Feudel}},\
  }\bibfield  {title} {\enquote {\bibinfo {title} {Numerical modelling of the
  effect of intermittent upwelling events on plankton blooms},}\ }\href@noop {}
  {\bibfield  {journal} {\bibinfo  {journal} {J. R. Soc. Interface}\ }\textbf
  {\bibinfo {volume} {17}},\ \bibinfo {pages} {20190889} (\bibinfo {year}
  {2020})}\BibitemShut {NoStop}%
\bibitem [{\citenamefont {Freilich}\ \emph {et~al.}(2022)\citenamefont
  {Freilich}, \citenamefont {Flierl},\ and\ \citenamefont
  {Mahadevan}}]{FFM2022}%
  \BibitemOpen
  \bibfield  {author} {\bibinfo {author} {\bibfnamefont {M}~\bibnamefont
  {Freilich}}, \bibinfo {author} {\bibfnamefont {G}~\bibnamefont {Flierl}}, \
  and\ \bibinfo {author} {\bibfnamefont {A}~\bibnamefont {Mahadevan}},\
  }\bibfield  {title} {\enquote {\bibinfo {title} {Diversity of growth rates
  maximizes phytoplankton productivity in an eddying ocean},}\ }\href@noop {}
  {\bibfield  {journal} {\bibinfo  {journal} {Geophys. Res. Lett.}\ }\textbf
  {\bibinfo {volume} {49}},\ \bibinfo {pages} {e2021GL096180} (\bibinfo {year}
  {2022})}\BibitemShut {NoStop}%
\bibitem [{\citenamefont {Craik}\ and\ \citenamefont
  {Leibovich}(1976)}]{CL1976}%
  \BibitemOpen
  \bibfield  {author} {\bibinfo {author} {\bibfnamefont {A~D~D}\ \bibnamefont
  {Craik}}\ and\ \bibinfo {author} {\bibfnamefont {S}~\bibnamefont
  {Leibovich}},\ }\bibfield  {title} {\enquote {\bibinfo {title} {A rational
  model for {Langmuir} circulations},}\ }\href@noop {} {\bibfield  {journal}
  {\bibinfo  {journal} {J. Fluid Mech.}\ }\textbf {\bibinfo {volume} {73}},\
  \bibinfo {pages} {401--426} (\bibinfo {year} {1976})}\BibitemShut {NoStop}%
\bibitem [{\citenamefont {Thorpe}(2004)}]{thorpe2004}%
  \BibitemOpen
  \bibfield  {author} {\bibinfo {author} {\bibfnamefont {S~A}\ \bibnamefont
  {Thorpe}},\ }\bibfield  {title} {\enquote {\bibinfo {title} {Langmuir
  circulation},}\ }\href@noop {} {\bibfield  {journal} {\bibinfo  {journal}
  {Annu. Rev. Fluid Mech.}\ }\textbf {\bibinfo {volume} {36}},\ \bibinfo
  {pages} {55--79} (\bibinfo {year} {2004})}\BibitemShut {NoStop}%
\bibitem [{\citenamefont {Moroz}\ and\ \citenamefont
  {Leibovich}(1985)}]{ML1985}%
  \BibitemOpen
  \bibfield  {author} {\bibinfo {author} {\bibfnamefont {I~M}\ \bibnamefont
  {Moroz}}\ and\ \bibinfo {author} {\bibfnamefont {S}~\bibnamefont
  {Leibovich}},\ }\bibfield  {title} {\enquote {\bibinfo {title} {Competing
  instabilities in a nonlinear model of {Langmuir} circulations},}\ }\href@noop
  {} {\bibfield  {journal} {\bibinfo  {journal} {Phys. Fluids}\ }\textbf
  {\bibinfo {volume} {28}},\ \bibinfo {pages} {2050--2061} (\bibinfo {year}
  {1985})}\BibitemShut {NoStop}%
\bibitem [{\citenamefont {Bees}(1998)}]{bees1998}%
  \BibitemOpen
  \bibfield  {author} {\bibinfo {author} {\bibfnamefont {M~A}\ \bibnamefont
  {Bees}},\ }\bibfield  {title} {\enquote {\bibinfo {title} {Planktonic
  communities and chaotic advection in dynamical models of {Langmuir}
  circulation},}\ }\href@noop {} {\bibfield  {journal} {\bibinfo  {journal}
  {Appl, Sci. Res.}\ }\textbf {\bibinfo {volume} {59}},\ \bibinfo {pages}
  {141--158} (\bibinfo {year} {1998})}\BibitemShut {NoStop}%
\bibitem [{\citenamefont {Taylor}\ and\ \citenamefont
  {Ferrari}(2011)}]{TF2011}%
  \BibitemOpen
  \bibfield  {author} {\bibinfo {author} {\bibfnamefont {J~R}\ \bibnamefont
  {Taylor}}\ and\ \bibinfo {author} {\bibfnamefont {R}~\bibnamefont
  {Ferrari}},\ }\bibfield  {title} {\enquote {\bibinfo {title} {Shutdown of
  turbulent convection as a new criterion for the onset of spring phytoplankton
  blooms},}\ }\href@noop {} {\bibfield  {journal} {\bibinfo  {journal} {Limnol.
  Oceanogr.}\ }\textbf {\bibinfo {volume} {56}},\ \bibinfo {pages} {2293}
  (\bibinfo {year} {2011})}\BibitemShut {NoStop}%
\bibitem [{\citenamefont {Lindemann}\ \emph {et~al.}(2017)\citenamefont
  {Lindemann}, \citenamefont {Visser},\ and\ \citenamefont
  {Mariani}}]{LVM2017}%
  \BibitemOpen
  \bibfield  {author} {\bibinfo {author} {\bibfnamefont {C}~\bibnamefont
  {Lindemann}}, \bibinfo {author} {\bibfnamefont {A}~\bibnamefont {Visser}}, \
  and\ \bibinfo {author} {\bibfnamefont {P}~\bibnamefont {Mariani}},\
  }\bibfield  {title} {\enquote {\bibinfo {title} {Dynamics of phytoplankton
  blooms in turbulent vortex cells},}\ }\href@noop {} {\bibfield  {journal}
  {\bibinfo  {journal} {J. R. Soc. Interface}\ }\textbf {\bibinfo {volume}
  {14}},\ \bibinfo {pages} {20170453} (\bibinfo {year} {2017})}\BibitemShut
  {NoStop}%
\bibitem [{\citenamefont {Tergolina}\ \emph {et~al.}(2021)\citenamefont
  {Tergolina}, \citenamefont {Calzavarini}, \citenamefont {Mompean},\ and\
  \citenamefont {Berti}}]{TCMB2021}%
  \BibitemOpen
  \bibfield  {author} {\bibinfo {author} {\bibfnamefont {V.B.}\ \bibnamefont
  {Tergolina}}, \bibinfo {author} {\bibfnamefont {E.}~\bibnamefont
  {Calzavarini}}, \bibinfo {author} {\bibfnamefont {G.}~\bibnamefont
  {Mompean}}, \ and\ \bibinfo {author} {\bibfnamefont {S.}~\bibnamefont
  {Berti}},\ }\bibfield  {title} {\enquote {\bibinfo {title} {Effects of
  large-scale advection and small-scale turbulent diffusion on vertical
  phytoplankton dynamics},}\ }\href@noop {} {\bibfield  {journal} {\bibinfo
  {journal} {Phys. Rev. E}\ }\textbf {\bibinfo {volume} {104}},\ \bibinfo
  {pages} {065106} (\bibinfo {year} {2021})}\BibitemShut {NoStop}%
\end{thebibliography}%

\end{document}